\documentclass[12pt,showpacs,aps]{revtex4}

\usepackage{amsmath}
\usepackage{amssymb}
\usepackage{dsfont} 

\usepackage{graphicx}

\usepackage{color}

\newcommand{\be}{\begin{equation}}
\newcommand{\ee}{\end{equation}}
\newcommand{\bel}[1]{\begin{equation}\label{#1}}
\newcommand{\bea}{\begin{eqnarray}}
\newcommand{\eea}{\end{eqnarray}}
\newcommand{\balign}{\begin{align}}
\newcommand{\ealign}{\end{align}}
\newcommand{\ba}{\begin{array}}
\newcommand{\ea}{\end{array}}
\newcommand{\bfig}{\begin{figure}}
\newcommand{\efig}{\end{figure}}

\newcommand{\eref}[1]{(\ref{#1})}

\newcommand{\Sref}[1]{Section~\ref{#1}}
\newcommand{\Aref}[1]{Appendix~\ref{#1}}
\newcommand{\Fref}[1]{Fig.~\ref{#1}}

\allowdisplaybreaks

\newcommand{\exval}[1]{\mbox{$\langle \, {#1}\, \rangle$}}

\newcommand{\rmd}{\mathrm{d}}
\newcommand{\rme}{\mathrm{e}}

\newcommand{\sign}{\mathrm{sgn}}

\newcommand{\pdt}{\frac{\partial}{\partial t}}
\newcommand{\pdx}{\frac{\partial}{\partial x}}
\newcommand{\half}{\frac{1}{2}}

\begin{document}

\title{Universality classes in two-component driven diffusive systems}
\author{V. Popkov$^{(1,2)}$, J. Schmidt$^{(1)}$}
\affiliation{$^{(1)}$Institut f\"{u}r Theoretische Physik, Universit\"{a}t zu K\"{o}ln, Z\"ulpicher Str. 77,
50937 Cologne, Germany.}
\affiliation{$^{(2)}$CSDC Universit\`a di Firenze, via G.Sansone 1, 50019 Sesto Fiorentino, Italy}
\author{G.M.~Sch\"utz$^{(3,4)}$}
\affiliation{$^{(3)}$Institute of Complex Systems II, Theoretical Soft Matter and Biophysics,
Forschungszentrum J\"ulich, 52425 J\"ulich, Germany}
\affiliation{$^{(4)}$Interdisziplin\"ares Zentrum f\"ur Komplexe Systeme, Universit\"at
Bonn, Br\"uhler Str. 7, 53119 Bonn, Germany}

\begin{abstract}
We study time-dependent density fluctuations in the stationary state of driven diffusive systems
with two conserved densities $\rho_\lambda$. Using Monte-Carlo simulations
of two coupled single-lane asymmetric simple exclusion processes we present numerical evidence
for universality classes with dynamical exponents $z=(1+\sqrt{5})/2$ and $z=3/2$
(but different from the Kardar-Parisi-Zhang (KPZ) universality class), which have not been reported
yet for driven diffusive systems.
The numerical asymmetry of the dynamical structure functions converges slowly for some of the
non-KPZ superdiffusive modes for which mode coupling theory predicts
maximally asymmetric $z$-stable L\'evy scaling functions. We show that all universality 
classes predicted by mode coupling theory for two conservation laws are generic:
They occur in two-component systems with 
nonlinearities in the associated currents already of the minimal order $\rho_\lambda^2\rho_\mu$.
The macroscopic stationary current-density relation and the compressibility matrix 
determine completely all permissible universality classes through the 
mode coupling coefficients which we compute explicitly for general two-component systems.

\end{abstract}

\date{\today }

\maketitle


\newpage

\section{Introduction}
\label{s_intro}

Anomalous transport is the hallmark of many one-dimensional non-equilibrium systems
even when interactions are short-ranged \cite{Lepr03}.
A common way of characterizing 1-d systems that exhibit anomalous transport
is through the dynamical structure function which describes the time-dependent
fluctuations of the long-lived modes in the stationary state. In systems with
short-range interactions and one global conservation law (giving rise to one long-lived mode)
only two universality classes are known to exist, the Gaussian universality
class with dynamical exponent $z=2$ (also describing diffusive fluctuations in equilibrium stationary
states), and the superdiffusive Kardar-Parisi-Zhang (KPZ) universality class with dynamical
exponent $z=3/2$ \cite{Kard86} for systems driven out of equilibrium. The exact scaling form of the
KPZ structure function was found
some 10 years ago by Pr\"ahofer and Spohn for the polynuclear growth model \cite{Prae04}
and for a driven diffusive system, viz. the asymmetric simple exclusion process \cite{Prae02}.
Since then the scaling function, which is expected to be universal, has also been
observed in various experiments \cite{Miet05,Take10}.

Superdiffusive fluctuations in systems with more than one conservation law are less well-studied.
Stochastic dynamics have been considered for driven diffusive systems with two conservation laws.
Naively one might expect both modes to be in the KPZ universality class. This guess is indeed confirmed
for the Arndt-Heinzel-Rittenberg model \cite{Arnd98b} by using exact results for the steady state combined
by fluctuating hydrodynamics and mode coupling theory \cite{Ferr13} and also
for a general class of multi-component exclusion processes \cite{Kauf15}. It was also known for some time
that one mode can be KPZ, while the other is diffusive, see \cite{Rako04} where exact microscopic and
hydrodynamic limit arguments are used, and numerical work \cite{Das01,Naga05} for related results.

Recently van Beijeren \cite{vanB12} studied a system with Hamiltonian dynamics with three
conservation laws. He predicted KPZ-universality for the two sound modes of the system and a
novel superdiffusive universality class with dynamical exponent $z=5/3$ for the heat mode.
The occurrence of a 5/3 mode was subsequently demonstrated for FPU-chains \cite{Mend13,Das14}
with three conservation laws and generally for anharmonic chains \cite{Spoh14} and a family of
exclusion process with two conservation laws \cite{Popk14}. Also recent mathematically rigorous
work indicates non-trivial anomalous behaviour fluctuations in systems with two
conservation laws \cite{Bern14}.

Stochastic interacting particle systems with two conservation laws exhibit extremely rich
behaviour in one dimension, including spontaneous symmetry breaking
\cite{Arnd98b,Evan95,Popk01,Will05,Popk08,Gupt09} or phase separation
\cite{Arnd98b,Popk01,Evan98b,Lahi00,Mett02,Clin02} in nonequilibrium stationary states,
see \cite{Schu03} for a review.
Studying the coarse-grained time evolution of two-component systems
with an umbilic point one finds shocks with unusual properties \cite{Popk12,Popk13}.
It is the purpose
of this paper to go beyond stationary and time-dependent mean properties
and consider time-dependent fluctuations. Specifically,
we show that the complete list of dynamical universality classes
that, according to mode coupling theory, can appear in the presence
of two conservation laws can be realized
in driven diffusive systems with two conserved densities.
To this end we compute the exact mode coupling
matrices for general strictly hyperbolic two-component systems with the stationary
current-density relation and stationary compressibility matrix as the only input.
With these input data the scaling form of the dynamical structure function is
completely determined, except in the presence of a diffusive
mode where the phenomenological diffusion coefficient enters the scale factors
in the scaling functions. With these results we use mode coupling theory
for computing explicitly the scaling form of the dynamical structure function for
two superdiffusive modes which have been not reported yet in the literature on
driven diffusive systems. We also present simulation data for a family of exclusion processes
which confirm the theoretical predictions.

This paper is organized in the following way. We first introduce the lattice model that we are going
to study numerically (\Sref{Sec:Model}). This is an extended version of the two-lane exclusion process
presented in our earlier work \cite{Popk14} that allows us to relax constraints on the physically accessible
parameter manifold. In \Sref{Sec:MCT} we first present some predictions of mode coupling theory
and then use the theory to make predictions for our model. The numerical tests
of these predictions and some mode coupling computations
are presented in \Sref{Sec:Simulation}. We finish with some conclusions in
\Sref{Sec:Conclusions}. In the appendix we perform the full computation of  the mode coupling
matrices for arbitrary strictly hyperbolic two-component systems.

\section{Two-lane asymmetric simple exclusion process}
\label{Sec:Model}

We consider a two-lane asymmetric simple exclusion process where particles hop randomly
on two parallel chains with $L$ sites each and periodic boundary conditions. Particles
do not change lanes and they obey the hard core exclusion principle which forbids
occupancy of a site by more than one particle.
We denote the particle occupation number on site $k$ in
the first (upper) lane by $n^{(1)}_{k}\in \{0,1\}$ , and on the second (lower) lane by
$n^{(2)}_{k}\in \{0,1\}$. The total particle number is conserved in each lane and denoted
$N_\lambda$.

\begin{figure}[ptb]
\begin{center}
\includegraphics[scale=0.6]{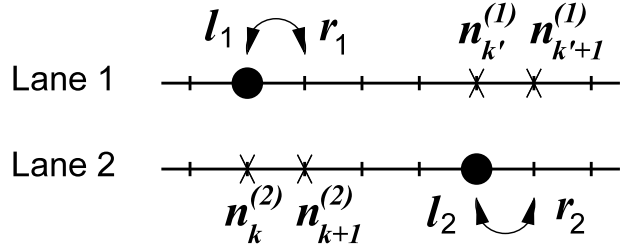}
\end{center}
\caption{Schematic representation of the two-lane partially asymmetric simple exclusion process.
A particle on lane 1 (2) hops to the neighbouring site (provided this target site is empty) with
to the right or left with rates \eref{hoppingrates} that depend on the
particle configuration on the adjacent sites of the other lane that are marked by a cross.}
\label{Fig:model}
\end{figure}

A hopping event from site $k$ to site $k+1$ on the same lane may happen if site
$k$ is occupied and site $k+1$ on the same lane is empty. The rate of hopping
depends on the particle configuration on the adjacent lane as follows:
Particles on lane $\lambda$ hop from site $k$ to site $k+1$ with rate $r_\lambda(k,k+1)$
and from site $k+1$ to site $k$ with rate $\ell_\lambda(k+1,k)$  (\Fref{Fig:model}). The rates
are given by
\be
\label{hoppingrates}
\ba{lll}
r_1(k,k+1) & = &  p_1 + b_1 n^{(2)}_k + c_1 n^{(2)}_{k+1} + d_1 n^{(2)}_k n^{(2)}_{k+1}  \\
\ell_1(k+1,k) & = & q_1 + e_1 n^{(2)}_k + f_1 n^{(2)}_{k+1} + g_1 n^{(2)}_k n^{(2)}_{k+1}  \\
r_2(k,k+1) & = & p_2 + b_2 n^{(1)}_k + c_2 n^{(1)}_{k+1} + d_2 n^{(1)}_k n^{(1)}_{k+1}  \\
\ell_2(k+1,k) & = &  q_2 + e_2 n^{(1)}_k + f_2 n^{(1)}_{k+1} + g_2 n^{(1)}_k n^{(1)}_{k+1} .
\ea
\ee
The hopping attempts of particles from site $k$ on lane $\lambda$ to neighbouring sites
occur independently of each other, after
an exponentially distributed random time with  mean $\tau_\lambda(k) =
[r_\lambda(k,k+1) + \ell_\lambda(k,k-1)]^{-1}$ for a jump from site $k$ on lane $\lambda$.
Hopping attempts on an already occupied site are rejected.

Using pairwise balance \cite{Schu96} it is easy to verify that for any pair of total particle
numbers $N_\lambda$ the stationary distribution
for this model is the uniform distribution, provided
that the symmetry constraints $b_1-e_1=c_2-f_2$, $b_2-e_2=c_1-f_1$, $d_1=g_1$ and $d_2=g_2$
are met for the interaction constants between the two lanes.
The ``bare'' hopping rates $p_1,p_2,q_1,q_2$ are arbitrary.
From the canonical uniform measures one constructs stationary grandcanonical
product measures where each site of lane $\lambda$ is occupied independently of the other sites
with probability $\rho_\lambda \in [0,1] = N_\lambda/L$. Hence the $\rho_\lambda$ are the conserved densities
of the grandcanonical stationary distribution, which, by construction, is the convex combination
of all uniform measures with weight $[\rho_1/(1-\rho_1 )]^{N_1} [\rho_2/(1-\rho_2 )]^{N_2}$
and $0\leq N_\lambda \leq L$.

From the hopping rates \eref{hoppingrates} and the product form of the grandcanonical 
distribution one reads off the
corresponding stationary current vector $\vec{j}$ with components
\be
\label{current}
\ba{lll}
j_1 (\rho_1,\rho_2) & = & \rho_1(1-\rho_1)(a + \gamma \rho_2), \\
j_2 (\rho_1,\rho_2) & = & \rho_2(1-\rho_2)(b + \gamma \rho_1).
\ea
\ee
with
\be
a=p_1-q_1,\, b=p_2-q_2,\, \gamma = b_1+c_1-e_1-f_1.
\ee
Notice that this current-density relation depends on the
microscopic details of the model only through the parameter combinations
$a,b,\gamma$ which can take arbitrary real values. 
For $b=1$ we recover the totally asymmetric two-lane model of \cite{Popk03}
which is a special case of the multi-lane model of \cite{Popk04}.
Throughout this work we set $a=1$, $\gamma \neq 0$.

The product measure corresponds to a grandcanonical ensemble
with a fluctuating particle number. These fluctuations are described by the
symmetric compressibility matrix $K$ with matrix elements
\begin{equation}
\label{K}
K_{\lambda\mu} := \frac{1}{L}
\left< (N_\lambda - \rho_\lambda L)(N_\mu - \rho_\mu L) \right> = 
\rho_\lambda(1-\rho_\lambda)\delta_{\lambda,\mu} .
\end{equation}
where $\lambda, \mu \in \{1,2\}$. In the notation defined in the appendix this
corresponds to
\bel{compressibilities}
\kappa_\lambda := K_{\lambda\lambda} = \rho_\lambda(1-\rho_\lambda), \quad \bar{\kappa} := K_{12} = 0.
\ee
As discussed below the current density relation $\vec{j}$ given in \eref{current}
and the compressibility matrix $K$ given \eref{K} are the input data
which completely determine the scaling functions describing the large scale behaviour of the
particle system, up to a scale factor if a diffusive mode is relevant.

For the Monte-Carlo simulations presented in this paper we consider the 
totally asymmetric version of the model \cite{Popk14} where $p_1=1$, $p_2=b$,
$q_\lambda=e_\lambda=f_\lambda=g_\lambda=d_\lambda=0$
and $b_\lambda=c_\lambda=\gamma/2 \neq 0$ with $\gamma > - \min{(1,b)}$. 
Initially we put $N_\lambda$ particles randomly drawn from the stationary distribution, 
i.e., they are placed
uniformly on lane $\lambda$. For the dynamics we perform random sequential updates
where a site $k_\lambda$ is chosen uniformly and a particle, if present 
and allowed to jump, jumps 
with a normalized probability given by \eref{hoppingrates}. One Monte-Carlo time 
unit then corresponds to $2L$ consecutive update attempts.
We compute the empirical dynamical structure function defined by
$\bar{S}_{k}^{\lambda\mu}(t)=1/n \sum_{j=1}^{n} 1/L 
\sum_{l=1}^L n_{l+k}^{(\lambda)}(j\tau+t)n_{l}^{(\mu)}(j\tau) - \rho_\lambda\rho_\mu$
where for numerical efficiency we exploit translation variance and 
take a sum over $n$ multiples of $\tau$ and over $m$ Monte-Carlo histories.
Time $t$ and system size $L$ are chosen such that finite-size corrections to the
stationary current (which are of order $1/L$) and to the structure function (at most of order
$1/L^{1+\alpha}$ with $\alpha>1$ as discussed below) are small in absolute terms
and negligible compared to statistical errors.

\section{Dynamical universality classes}
\label{Sec:MCT}

\subsection{Fluctuating hydrodynamics and mode coupling theory}

Following the ideas set out in \cite{Spoh91,Kipn99} the starting point for investigating the
large-scale dynamics of a microscopic lattice model is the
system of conservation laws
\bel{hydro}
\pdt \vec{\rho}(x,t) + \pdx \vec{j}(x,t) = 0
\ee
where component $\rho_\lambda(x,t)$ of the density vector $\vec{\rho}(x,t)$ is the 
coarse-grained local density
of the component $\lambda$ of the system, and the component $j_\lambda(x,t)$ 
of the current vector $\vec{j}(x,t)$ is the
associated current. The current is a function of $x$ and $t$ only through its dependence on the local
conserved densities.
Hence these equations can be rewritten as
\begin{equation}
\label{hyper}
\frac{\partial}{\partial t} \vec{\rho}(x,t) + J \frac{\partial}{\partial x} \vec{\rho}(x,t) = 0
\end{equation}
where $J$ is the
current Jacobian with matrix elements $ J_{\lambda\mu} = \partial j_\lambda / \partial \rho_\mu$.
The product $JK$ of the Jacobian with the compressibility matrix \eref{K} is
symmetric \cite{Gris11} which
guarantees that the system (\ref{hyper}) is hyperbolic \cite{Toth03}.
The eigenvalues $v_\alpha$ of $J$ are the characteristic velocities of the system.
If $v_1 \neq v_2$ the system is called strictly hyperbolic.
Notice that in our convention $\vec{\rho}$ and  $\vec{j}$ are regarded as column vectors.
Transposition is denoted by a superscript $T$.

Eq. (\ref{hyper}) describes the deterministic time evolution of the density under Eulerian scaling
where the lattice spacing $a$ is taken to zero
such that $x=ka$ remains finite and at the same time the microscopic time $\tau$ is taken to infinity
such that the macroscopic time $t=\tau a$ is finite.
The effect of fluctuations, which occur on finer space-time scales
where $t=\tau a^z$ with dynamical exponent $z>1$, can be
captured by adding phenomenological white noise terms $\xi_i$
and taking the non-linear
fluctuating hydrodynamics approach together with a mode-coupling analysis of the non-linear
equation. Following \cite{Spoh14}
we summarize here the main ingredients of this well-established description.

One expands the local densities $\rho_\lambda(x,t) = \rho_\lambda + u_\lambda(x,t) $
around their long-time stationary values $\rho_\lambda$
and keeps terms to first non-linear order in the
fluctuation fields $u_\lambda(x,t)$.
For quadratic nonlinearities \eref{hyper} then yields
\begin{equation}
\label{coupledBurgers}
\partial_t \vec{u} =  - \partial_x \left( J \vec{u} + \frac{1}{2} \vec{u}^T \vec{H} \vec{u} - D \partial_x  \vec{u}
+ B \vec{\xi} \right)
\end{equation}
where $\vec{H}$ is a column vector whose entries $(\vec{H})_\lambda=H^{\lambda}$ are the Hessians
with matrix
elements $H^{\lambda}_{\mu\nu} = \partial^2 j_\lambda /(\partial \rho_\mu \partial \rho_\nu)$.
The term $\vec{u}^T H^{\lambda} \vec{u}$ denotes the inner product in component space.
The diffusion matrix $D$ is a phenomenological quantity. The noise strength $B$ does
not appear explicitly below, but plays an indirect role in the mode-coupling analysis.
One recognizes in \eref{coupledBurgers} a system of coupled
noisy Burgers equations.
If the quadratic non-linearity is absent one has diffusive behaviour,
up to possible logarithmic corrections that may arise from cubic non-linearities \cite{Devi92}.

In order to analyze this nonlinear equation we transform to normal modes $\vec{\phi} =R \vec{u}$
where $ RJR^{-1} = \mathrm{diag}(v_\alpha)$ and the transformation matrix $R$ is normalized
such that $RKR^T = \mathds{1}$, see the appendix. From \eref{coupledBurgers} one thus arrives at
\begin{equation}
\label{normalmodes}
\partial_t \phi_\alpha = -  \partial_x \left( v_\alpha \phi_\alpha +
\vec{\phi}^T G^{\alpha} \vec{\phi} - \partial_x  (\tilde{D} \vec{\phi})_\alpha
+ (\tilde{B} \vec{\xi})_\alpha \right)
\end{equation}
with $\tilde{D}=RDR^{-1}$, $\tilde{B}=RB$ and
\begin{equation}
G^{\alpha} =  \frac{1}{2} \sum_\lambda R_{\alpha\lambda} (R^{-1})^T H^{\lambda} R^{-1}
\end{equation}
are the mode coupling matrices.

To make contact of this macroscopic description
with the microscopic model we first note that the current-density relation
given by the components of the current vector $\vec{j}$
arises from the microscopic model by computing the stationary current-density
relations $j_\lambda(\rho_1,\rho_2)$ and then substituting the stationary conserved densities
by the coarse-grained local densities $\rho_\lambda(x,t)$ which are regarded as
slow variables. Similarly, the compressibility matrix $K$ is
computed from the stationary distribution. Hence the mode coupling matrices (and with them
the dynamical universality classes as shown below) are
completely determined by these two macroscopic stationary properties of the system.
We stress that the {\it exact} stationary current-density relations and
the {\it exact} stationary compressibilities are required. Approximations obtained e.g.
from stationary mean field theory will, in general, only accidentally
provide the information necessary for determining the dynamical universality classes
of the system.
In the appendix we compute the mode coupling matrices of a general two-component system
with the current vector and compressibility matrix as input parameters.

Second, consider the dynamical structure matrix $\bar{S}_k(t)$ of the microscopic model
defined on the lattice.\footnote{We choose the same notation as for the empirical
structure function obtained from Monte-Carlo simulations presented below.} Its matrix elements
are the dynamical structure functions
\bel{dynstrucfun}
\bar{S}_k^{\lambda\mu}(t) : = \exval{(n^{(\lambda)}_k(t) - \rho_\lambda)(n^{(\mu)}_0(t) - \rho_\mu)}
\ee
which measure density fluctuations in the stationary state.
This quantity has two different physical interpretations. On the one hand, one can
regard the random variable
$f^{\lambda}_k(t) := n^{(\lambda)}_k(t) - \rho_\lambda$
as a stochastic process and then the dynamical structure
function describes the stationary two-time correlations of this process.
The long-time behaviour of the dynamical structure function
can thus be determined from the fluctuation fields $u_\lambda(x,t)$ appearing in the
non-linear fluctuating hydrodynamics approach \eref{coupledBurgers}, i.e.,
$\bar{S}_k^{\lambda\mu}(t) \stackrel{k,t\to\infty}{\to} \exval{u_\lambda(x,t)u_\mu(0,0)}$.
In a different interpretation the dynamical structure
function measures the time evolution of the expectation of $f^{\lambda}_k(t)$ at time $t$,
i.e., the unnormalized density
profiles $\bar{\rho}^{\lambda}_k(t) := \exval{n^{(\lambda)}_k(t) - \rho_\lambda}$ that at time $t=0$
have a delta-peak at site 0.
Since the two conserved quantities
interact, an initial perturbation even of only one component  will cause a non-trivial
relaxation of {\it both} density profiles. In each component the initial peak will evolve into
two separate peaks, which move and spread with time.
The characteristic velocities $v_\alpha$ are the collective
velocities, i.e., the center-of-mass
velocities of the two local perturbations \cite{Popk03}. The variance of the evolving
density profiles determines the collective diffusion coefficient.
This second interpretation of the dynamical structure matrix as describing a relaxation process,
completely equivalent to the first fluctuation interpretation,
is quite natural
from the viewpoint of regarding \eref{coupledBurgers} as a more detailed description
of \eref{hydro} in the sense of describing fluctuation effects on finer space-time scales due to
the randomness of the stochastic process from which \eref{hydro} arises under Eulerian scaling.

Analogously one can regard the transformed modes of the lattice model
$\vec{\phi}_k (t) = R \vec{f}_k(t)$ in the fluctuation interpretation as
stationary processes and the
transformed dynamical structure functions
\bel{S-matrix}
S^{\alpha\beta}_k(t)=[R\bar{S}_k(t)R^T ]_{\alpha\beta} = \exval{\phi_k^{\alpha}(t)\phi_0^{\beta}(0)}
\ee
as the stationary space-time fluctuations.
The transformation of the dynamical structure functions
to the normal modes $\vec{\phi}_k (t)$ on the lattice,
which is important for the numerical simulation of lattice models,
is discussed in more detail in \Aref{Normalmodesbasis}. The large-scale behaviour of $S^{\alpha\beta}_k(t)$
is given in terms of the
normal modes $\phi_\alpha(x,t)$ appearing in \eref{normalmodes}
by $S_{\alpha\beta}(x,t) = \exval{\phi_{\alpha}(x,t)\phi_{\beta}(0,0)}$.
In the second relaxation interpretation the normal modes are seen as local perturbations of a stationary
distribution with a specific choice of initial amplitudes in each component.

Since for
strictly hyperbolic systems the two characteristic velocities are different, one
expects that the off-diagonal elements of $S$ decay quickly. For long times
and large distances one is thus left with the diagonal elements
which we denote by
\bel{S-diag}
S_\alpha(x,t):=S^{\alpha\alpha}(x,t)
\ee
with initial value $S_\alpha(x,0)=\delta(x)$.
The large scale behaviour of the diagonal elements is expected to have the scaling form
\bel{scalingform}
S_\alpha(x,t) \sim t^{-1/z_\alpha} f_{\alpha} ((x-v_\alpha t)^{z_\alpha}/t)
\ee
with a dynamical exponent $z_\alpha$ that may be different for the two modes. The exponent
in the power law prefactor follows from mass conservation.
In momentum space one has
\bel{scalingformFT}
\hat{S}_\alpha(p,t) \sim  \rme^{- iv_\alpha pt} \hat{f}_{\alpha} (p^{z_\alpha}t)
\ee
for the Fourier transform
\bel{Def:FT}
\hat{S}_\alpha(p,t) := \frac{1}{\sqrt{2\pi}}\int_{-\infty}^\infty \rmd x\, \rme^{- i px} S_\alpha(x,t).
\ee

Whether the difference of the characteristic speeds vanishes or not plays an important role.
For the case where $v_1=v_2$, i.e., when the system \eref{hyper} has an
umbilic point, it was found numerically in the framework of dynamic roughening of directed lines
that the dynamical exponent is $z=3/2$, but the scaling functions are not KPZ \cite{Erta92}.
On the other hand, for strictly hyperbolic systems the normal modes have different speeds
and hence their interaction
becomes very weak for long times. By identifying $\phi_\alpha$ with the gradient of a height variable
\eref{normalmodes} then turns generically into two decoupled KPZ-equations
with coefficients $G^{\alpha}_{\alpha\alpha}$ determining the strength of the nonlinearity.

In order to analyze the system of nonlinear stochastic PDE's in more detail
we employ mode coupling theory \cite{Spoh14}.
The basic idea is to capture the combined effect of non-linearity and noise by a memory
kernel. Thus the starting point for computing the $S_\alpha(x,t)$ are the mode coupling equations
\be
\label{modecoupling}
\partial_t S_\alpha(x,t) = (-v_\alpha \partial_x + D_\alpha \partial_x^2) S_\alpha(x,t)
+ \int_0^t \rmd s \int_{-\infty}^\infty \rmd y\, S_\alpha(x-y,t-s) \partial_y^2 M_{\alpha\alpha}(y,s)
\ee
with the diagonal element $D_\alpha:=\tilde{D}_{\alpha\alpha}$ of the phenomenological
diffusion matrix and the memory kernel
\bel{memorykernel}
M_{\alpha\alpha}(y,s) = 2 \sum_{\beta,\gamma} (G^{\alpha}_{\beta\gamma})^2 S_\beta(y,s)S_\gamma(y,s).
\ee
The strategy is to plug into this equation, or into its Fourier representation, the scaling ansatz
\eref{scalingform} (or \eref{scalingformFT}). One gets equations
for the dynamical exponents arising from requiring non-trivial scaling solutions
and using the known results $z=3/2$ for KPZ and $z=2$ for diffusion.
In a next step one can then solve for the actual scaling functions,
see below. Since for $v_1\neq v_2$ one has $S_\beta(y,s)S_\gamma(y,s)\approx 0$ for $\beta\neq \gamma$
it is clear that the scaling behaviour of the solutions of \eref{modecoupling} will be determined
largely by the diagonal terms $G^{\alpha}_{\beta\beta}$ of the mode coupling matrices $G^{\alpha}$.
If a leading self-coupling term $G^{\alpha}_{\alpha\alpha}$ vanishes, one finds non-KPZ
behaviour for mode $\alpha$. In particular, if all diagonal terms are zero, the mode is diffusive.
A coupling of a diffusive mode to a KPZ-mode leads to a modified KPZ-mode \cite{Spoh14b}.
Thus the crucial property of the mode coupling matrices is whether a diagonal element is zero or not.

Some algebra along the lines of \cite{Spoh14} involving power counting 
then yields the complete list of possible universal
classes of strictly hyperbolic two-component systems from the structure of the
mode coupling matrices $G^{\alpha}$ as shown in Table \ref{Tab:MCTscenarios},
see also \cite{Spoh14b} where a similar table was derived independently. 
The shorthand
KPZ represents the KPZ scaling function, while KPZ' refers to modified KPZ, 
both with dynamical exponent $z=3/2$.
$D$ represents a Gaussian scaling function $f_{\alpha}$ with dynamical exponent $z_\alpha=2$,
$z_\alpha$L represents a $z_\alpha$-stable L\'evy distribution as scaling function $f_{\alpha}$
with dynamical exponent $z_\alpha$, GM (for golden mean) represents $\varphi$L with
$\varphi = (1+\sqrt{5})/2$.
In what follows we apply these general results to the two-lane model defined above.
It will transpire that all theoretically possible scenarios can actually be realized in this
family of models.

\begin{table}[htb]
\begin{tabular}{l|cccc}
\hline\hline
$\ba{ll} \diagdown & G^{1} \\ G^{2} &  \diagdown \ea$  &
$\left(\ba{ll} \star & \\ & \bullet \ea \right)$ &
$\left(\ba{ll} \star & \\ & 0 \ea \right)$ &
$\left(\ba{ll} \mathbf{0} & \\ & \bullet \ea \right)$ &
$\left(\ba{ll} \mathbf{0} & \\ & 0 \ea \right)$ \vspace*{1mm} \\
 \hline
$\left(\ba{ll} \bullet & \\ & \star \ea \right)$
& (KPZ,KPZ) & (KPZ,KPZ) & ($\frac{5}{3}$L,KPZ) &(D,KPZ') \\
$\left(\ba{ll} 0 & \\ & \star \ea \right)$
& (KPZ,KPZ) & (KPZ,KPZ) & ($\frac{5}{3}$L,KPZ) &(D,KPZ) \\
$\left(\ba{ll} \bullet & \\ & \mathbf{0} \ea \right)$
& (KPZ,$\frac{5}{3}$L) & (KPZ,$\frac{5}{3}$L) & (GM,GM) & (D,$\frac{3}{2}$L) \\
$\left(\ba{ll} 0 & \\ & \mathbf{0} \ea \right)$
& (KPZ',D) & (KPZ,D) & ($\frac{3}{2}$L,D) & (D,D) \\
\hline\hline
\end{tabular}
\caption{Classification of universal behaviour of the two modes by the structure of the mode coupling matrices $G^{\alpha}$. The acronyms denote: KPZ: KPZ universality class (superdiffusive),
KPZ': modified KPZ universality class (superdiffusive),
D = Gaussian universality class (normal diffusion),
$z_\alpha$L: superdiffusive universality class with $z_\alpha$-stable L\'evy
scaling function and GM = $\varphi$L with the golden mean $\varphi=(1+\sqrt{5})/2$.
An bullet or star in the $G^{\alpha}$ denotes a non-zero entry, no entry
represents an arbitrary value (zero or non-zero). The selfcoupling terms $G^{\alpha}_{\alpha\alpha}$ are marked as
star or boldface 0, resp.
}
\label{Tab:MCTscenarios}
\end{table}

\subsection{Mode-coupling matrix for the two-lane model}

The input data are the current-density relation \eref{current}
and the compressibility matrix \eref{K}. From the current-density relation one computes the current
Jacobian and the Hessian, which are used together with the compressibility matrix to compute the
basis  for normal modes and finally the
mode coupling matrices, as shown in detail in the appendix in the general case.

For the present model we remark first that the currents \eref{current}
are at most quadratic in each density. Hence no logarithmic corrections
to diffusive behaviour are expected in the two-lane model defined above.
Second, as discussed in the appendix,
in any coupled two-component system a vanishing cross compressibility
$\bar{\kappa} =0$ (where $\lambda \neq \mu$)
implies that the cross derivatives $\partial j_\lambda / \partial \rho_\mu$ 
of the currents have to be non-zero except
when one of the two components is frozen, i.e.,  fully occupied or fully empty.

For our system the explicit form of $J$ is
\begin{equation}\label{J}
J=
\begin{pmatrix}
(1+\gamma\rho_{2})(1-2\rho_{1}) & \gamma\rho_{1}(1-\rho_{1})\\
\gamma\rho_{2}(1-\rho_{2}) & (b+\gamma\rho_{1})(1-2\rho_{2})
\end{pmatrix}.
\end{equation}
and the Hessians $H^{\lambda}$ are

\begin{equation}
\label{H12}
H^{1}=
\left( \ba{cc}
-2(1+\gamma\rho_{2}) & \gamma(1-2\rho_{1})\\
\gamma(1-2\rho_{1}) & 0
\ea \right), \quad
H^{2}=
\left( \ba{cc}
0 & \gamma (1-2\rho_{2}) \\
\gamma (1-2\rho_{2}) & -2( b+\gamma \rho_{1} )
\ea \right).
\end{equation}

The parameters convenient for theoretical analysis
are not the matrix elements of the current Jacobian
and the Hessians, but the parameters $u$, $\omega = \tan{\vartheta}$ \eref{def:phi_u} and the
transformed Hessian parameters \eref{g1}, \eref{g2} defined in the appendix to which we refer
for the derivation of the following results. Here we point out only the relevant features of
the quantities resulting from these lengthy but simple computations.

The collective velocities $v_{1,2}$ are given in \eref{charvel}.
Notice that $J_{12}  J_{21} = \gamma^2 \rho_{1}(1-\rho_{1})\rho_{2}(1-\rho_{2}) \geq 0$ in the
whole physical parameter regime of the model.
In fact, unless one of the lanes is frozen
we have the strict inequality $ J_{12}  J_{21} > 0$. The frozen case is of
no interest since then the dynamics in the non-frozen lane
reduce to the dynamics of a single exclusion process.
Hence we shall assume $ J_{12}  J_{21} > 0$ throughout this paper.
Therefore the discriminant of the characteristic
polynomial of $J$ \eref{def:discrim} is non-zero
which implies
that the model is strictly hyperbolic in the parameter domain of interest.

The transformation matrix $R$ involves normalization factors $z_\pm$ \eref{R}
and the parameters $u$ and $\omega=\tan{\vartheta}$ defined in
\eref{def:phi_u}. 
From \eref{J} we find
\be
\omega = \frac{1-b -(2+b\gamma)\rho_1 + 
(\gamma+2b)\rho_2}{2\gamma\sqrt{\rho_{1}(1-\rho_{1})\rho_{2}(1-\rho_{2})}}
\left(1 + \sqrt{1+ \frac{4\gamma^2(\rho_{1}(1-\rho_{1})\rho_{2}(1-\rho_{2})) }{(1-b -(2+b\gamma)\rho_1 
+ (\gamma+2b)\rho_2)^2} } \right)
\ee
and
\be
u = \sqrt{\frac{\rho_{1}(1-\rho_{1})}{\rho_{2}(1-\rho_{2})}}.
\ee
For $ J_{11}= J_{22}$ one has $\omega=1$.

From the Hessians
\eref{H12} one obtains the mode coupling parameters \eref{g1}, \eref{g2}
\be
g^{1}_1 = -2(1+\gamma\rho_{2}), \quad
g^{1}_2 = 0, \quad
\bar{g}^{1} = \gamma \sqrt{\frac{\rho_{2}(1-\rho_{2})}{\rho_{1}(1-\rho_{1})}} (1-2\rho_{1}),
\ee
and
\be
g^{2}_1 = 0,\quad
g^{2}_2 = -2\sqrt{\frac{\rho_{2}(1-\rho_{2})}{\rho_{1}(1-\rho_{1})}}(b+\gamma\rho_{1}), \quad
\bar{g}^{2} = \gamma(1-2\rho_{2}).
\ee

The compressibility matrix enters the mode coupling coefficients only
through the normalization factors $z_\pm$ for which we obtain from \eref{zpmspecial}
\be
z_\pm = 1/\sqrt{\kappa_1}\notin \{0,\pm \infty \}.
\ee
This yields the desired diagonal elements of the mode coupling matrices
\be
G^{\alpha}_{\beta\beta}(\omega) = A_0 D^{\alpha}_{\beta}(\omega)
\ee
with
\bea
\label{Ddiag11twolane}
D^{1}_1(\omega) & = & g^{1}_1 - 2 \bar{g}^{1} \omega +  2 \bar{g}^{2} \omega^2 - g^{2}_2 \omega^3 \\
\label{Ddiag12twolane}
D^{1}_2(\omega) & = &  \left(2 \bar{g}^{1} - g^{2}_2\right) \omega
+ \left(g^{1}_1 - 2 \bar{g}^{2}\right) \omega^2  \\
\label{Ddiag21twolane}
D^{2}_1(\omega) & = &  \left(g^{1}_1 - 2\bar{g}^{2}\right) \omega
+ \left(g^{2}_2 - 2 \bar{g}^{1}\right) \omega^2  \\
\label{Ddiag22twolane}
D^{2}_2(\omega) & = &  g^{2}_2 +  2 \bar{g}^{2} \omega  +  2 \bar{g}^{1} \omega^2 + g^{1}_1 \omega^3.
\eea
and
\be
A_0=\half \sqrt{\kappa_1} \cos^3(\vartheta) \neq 0.
\ee

As discussed in the appendix the vanishing cross-compressibility
$\bar{\kappa}=0$ of our model guarantees that $A_0\neq 0$.
Therefore a diagonal element $G^{\alpha}_{\beta\beta}$ of a mode coupling matrix vanishes if and only if the
polynomial $D^{\alpha}_{\beta}$ defined in \eref{Ddiag11twolane} - \eref{Ddiag22twolane} vanishes.
In order to see whether all scenarios listed in Table \ref{Tab:MCTscenarios} can be realized
by making the appropriate diagonal matrix elements zero
we study  all these cases. The relation between vanishing diagonal elements and
the universality class as well as the values of the dynamical exponents 
follows from straightforward power counting in the mode coupling equations 
derived in \cite{Spoh14}, see below for the two special cases we focus on in this work.\\

\noindent \underline{Purely diffusive case (D,D):}\\

\noindent First consider the purely diffusive case (D,D) for which mode coupling theory requires
$D^{1}_1=D^{1}_2=D^{2}_1=D^{2}_2=0$. Demanding that
$D^{1}_2=D^{2}_1=0$ leads to the constraints
$g^{1}_1 = 2\bar{g}^{2}$ and $g^{2}_2 = 2\bar{g}^{1}$.
In terms of the parameters $b,\gamma,\rho_\lambda$ this reads
$- \gamma = 1/(1-\rho_2) = b/(1-\rho_1)$. This is outside the physical parameter range
$\gamma > -\min{(1,b)}$ of the totally asymmetric model of \cite{Popk14}, but can be
realized in the general two-lane model defined in \Sref{Sec:Model}.
Plugging this condition into $D^{1}_1=D^{2}_2=0$ yields the further conditions
that $g^{1}_1 = g^{2}_2 = 0$, i.e., both Hessians must vanish. This
requires
\be
\rho_1=\rho_2=1/2, \quad b=1, \quad \gamma = -2.
\ee
The characteristic velocities
are then $v_{1,2} = \mp 1$. It is somewhat counterintuitive that
for these values one has $j_1=j_2=0$, i.e., the system appears to be 
macroscopically in equilibrium,
but the Gaussian mass fluctuations travel with non-zero velocities.
A simple parameter choice for this scenario is $p_1=p_2=1$,
$q_1=d_1=g_1=q_2=d_2=g_2=0$, $b_1=c_1=b_2=c_2=-1/2$, $e_1=f_1=e_2=f_2=1/2$.\\

\noindent \underline{Superdiffusive mixed cases (D,KPZ'), (D,KPZ), (D,$\frac{3}{2}$L), (KPZ,$\frac{5}{3}$L):}\\

\begin{figure}[ptb]
\begin{center}
\includegraphics[scale=0.4]{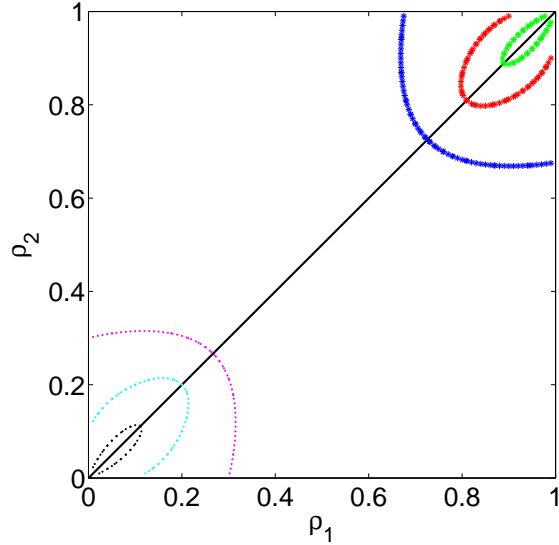}
\end{center}
\caption{(Colour online) Location of points where $G^{2}_{22}=0$, $G^{2}_{11} \neq 0$ for $b=1$ and different
values of $\gamma$. In the upper right (lower left) corner the points grouped along curves
of increasing length correspond to $\gamma=1.5,\,2.5,\,5$ ($\gamma = -0.6,-0.7,-0.85$).
On these curves one generically has the ($\frac{5}{3}$L,KPZ) universality class.
On the diagonal line $\rho_1=\rho_2$ one has $G^{1}_{11}=G^{1}_{22} = 0$, generically
corresponding to the (D,KPZ') universality class. On the intersection of this line
with a curve parametrized by $\gamma$ one has the (D,$\frac{3}{2}$L) universality class.
}
\label{Fig:b=1}
\end{figure}

\noindent
Consider $b=1$ where the hopping rates are completely symmetric with respect to the lane
interchange and take $\rho_1=\rho_2=:\rho$. Then $g^{1}_1 = g^{2}_2 = -2(1+\gamma\rho)$,
$g^{1}_2 = g^{2}_1 = 0$, $\bar{g}^{1}=\bar{g}^{1}=\gamma(1-2\rho)$ and $u=1$, $\omega = 1$.
This yields $D^{1}_1=D^{1}_2=0$ and $D^{2}_2=2 A_0 (g^{1}_1+ 2 \bar{g}^{1})$,
$D^{2}_1=2 A_0 (g^{1}_1 - 2 \bar{g}^{1})$ with $A_0 = \sqrt{\rho (1-\rho ) /32}$.
Computing the off-diagonal elements from \eref{G112}, \eref{G212} we find the full
mode coupling matrices
\begin{equation}
G^{1}= - 4 A_0 (1+ \gamma  \rho)
 \left(
\begin{array}{cc}
 0 & 1 \\
 1 & 0
\end{array}
\right), \quad
G^{2}=  - 4 A_0 \left(
\begin{array}{cc}
 1+\gamma  (1-\rho) & 0 \\
 0 &  1- \gamma  (1-3 \rho )
\end{array}
\right)
\end{equation}
Thus generically this line is in the (D,KPZ') universality class (\Fref{Fig:b=1}).

Notice that at $\gamma = - 1/(1-\rho)$ one has $D^{2}_1=0$, corresponding to the (D,KPZ)
universality class which can be realized in the generalized two-lane model defined above and
that occurs also in the single-lane multi-component asymmetric simple
exclusion process with stationary product measure \cite{Kauf15}. For
$\gamma = 1/(1-3\rho)$ one has
$D^{2}_2=0$, corresponding to the (D,$\frac{3}{2}$L) scenario, see next section.
If one moves away from the line $\rho_1=\rho_2$, but stays on the curves
indicated in \Fref{Fig:b=1} for special values of $\gamma$
the self-coupling coefficient $G^{1}_{11}$ is non-zero, but
$G^{2}_{22}=0$. This can be straightforwardly verified by calculating the linear response
of the diagonal elements of $G^1,G^2$ to  small deviations $\delta \rho_1,\delta \rho_2$ 
from the line $\rho_1=\rho_2$.
Hence one has the (KPZ,$\frac{5}{3}$L) scenario.
The three cases (D,KPZ'), (D,$\frac{3}{2}$L) and (KPZ,$\frac{5}{3}$L)
can be realized in the totally asymmetric two-lane model.\\

\noindent \underline{Golden mean universality class (GM,GM):}\\

\begin{figure}[ptb]
\begin{center}
\includegraphics[scale=0.4]{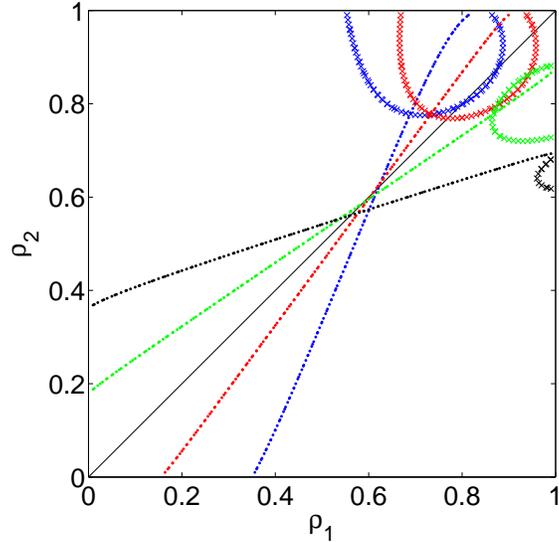}
\end{center}
\caption{(Colour online) Location of points where $G^{1}_{11} = 0$ $G^{1}_{22}\neq 0$ 
(crosses in the upper right corner)
or $G^{2}_{22}=0$ and $G^{2}_{11}\neq 0$ (thin bullets),
for fixed $\gamma = -3/4$ and $b = 1.5$ (black), $b = 1.2$ green), $b = 0.9$ (red),
$b = 0.8$ (blue), corresponding to the order from left to right in
the lower half of the figure and opposite order in the upper part of the figure.
Along the curves indicated by the dots (crosses) one has generically
the ($\frac{5}{3}$L,KPZ) or (KPZ,$\frac{5}{3}$L) universality class.
At the intersections of curves with the same colour one has the golden mean universality
class ($\varphi$L,$\varphi$L).
}
\label{Fig:bneq1}
\end{figure}

\noindent
Next consider $b \neq 1$. The formulas for the mode coupling matrices become cumbersome
and we do not present them here in explicit form in full generality. It turns out that one can have that
both self-coupling coefficients $G^{\alpha}_{\alpha\alpha}$ vanish and both subleading
diagonal elements $G^{\alpha}_{\beta\beta}$ with $\beta\neq \alpha$ 
are non-zero, corresponding to the ($\varphi$L,$\varphi$L)
scenario where both dynamical exponents are the golden mean $\varphi = (1+\sqrt{5})/2$,
see \Fref{Fig:bneq1}.
This can be realized by choosing unequal densities such that
\bel{GMparameters}
(1+\gamma\rho_{2})(1-2\rho_{1}) = (b+\gamma\rho_{1})(1-2\rho_{2})
\ee
which corresponds to $J_{11}=J_{22}$ and hence $\omega=1$.
Then the requirement $D^{1}_1=D^{2}_2=0$ yields
\bel{GMdensities}
\rho_1 = \frac{1-b}{3\gamma},\quad  \rho_2 = \frac{\gamma-1}{3\gamma}
\ee
which implies $\gamma \in (-\infty,-1/2] \cup [1,\infty)$ and $b$
is in the range between $\gamma$ and $-2\gamma$.
For general values of $\omega$ the analytical formulas for the
lines $G^{1}_{11}=G^{2}_{22}=0$ in the $\rho_1-\rho_2$ plane are complicated.
In order to demonstrate the existence of solutions we show numerical plots for
fixed $\gamma = -3/4$ and various values $b$ in \Fref{Fig:bneq1}. Notice also that
there are parameter ranges of $b$ without solutions in the physical range
of densities $(\rho_1,\rho_2) \in [0,1] \times [0,1]$.

In what follows we investigate in more detail the two novel universality classes
(D,$\frac{3}{2}$L) and (GM,GM)
which have not been reported yet in the
literature on driven diffusive systems.
We also comment on the shape of the structure function for the $\frac{5}{3}$-mode
discussed in \cite{Popk14}.

\section{Superdiffusive non-KPZ universality classes}
\label{Sec:Simulation}

\subsection{Diffusive mode and 3/2\,-\,L\'evy mode}

We consider the case where mode 1 is Gaussian, and mode 2 has non-vanishing cross-coupling,
\begin{equation}
G^{1}_{11}=G^{1}_{22}=G^{2}_{22}=0, \quad  G^{2}_{11} \neq 0
\label{New
Mode}
\end{equation}
The mode coupling equation \eref{modecoupling} for mode 2 reads in Fourier space
\bea
\partial_t \hat{S}_2(p,t) & = & -  i p v_2   \hat{S}_2(p,t) -  p^2 D_2  \hat{S}_2
 \nonumber \\
& & - 2 (G^{2}_{11})^2  p^2 \int_0^t ds  \hat{S}_2(p,t-s)
\int_{-\infty}^{\infty} dq  \hat{S}_1((p-q,s) \hat{S}_1((q,s).
\label{ModeCoupling2}
\eea
with $D_2=\tilde{D}_{22}$.
For the Gaussian mode 1 the mode coupling  equation is obtained by the
exchange $1\leftrightarrow 2$ in (\ref{ModeCoupling2})
and dropping the term containing the integral. Note that we are
interested in the large $x$ behaviour of the scaling function,
meaning $p \rightarrow 0$ in Fourier space.

We start with the observation that the Gaussian mode has the usual scaling form
\begin{equation}
S_1(x,t)=\frac{1}{ \sqrt{4 \pi D_1 t}} \rme^{-\frac{(x-v_1 t)^2}{4 D_1 t}}
\label{S2(x,t)}
\end{equation}
with Fourier transform
$\hat{S}_1(p,t) = 1/\sqrt{2\pi} \exp\left(- i v_1 p t - D_1 p^2 t\right)$.
Inserting this into (\ref{ModeCoupling2}) and performing the integration over $q$, we obtain
\be
\partial_t \hat{S}_2(p,t) = - ( i p v_2  +  p^2 D_2 )  \hat{S}_2(p,t)
 -  p^2 \left(G^{2}_{11}\right)^2  \int_0^t ds  \hat{S}_2(p,t-s)
 \frac{\rme^{- i v_1 p s - D_2 p^2 s/2}}{\sqrt{2 \pi  D_2 s}} .
\label{MC0}
\ee
This equation can be solved in terms of the Laplace transform
$\tilde{S}_2(p,\omega) := \int_0^\infty \rmd t \rme^{-\omega t} \hat{S}_2(p,t)$ which yields
\bel{LPMCO}
\tilde{S}_2(p,\omega) =
\frac{\hat{S}_2(p,0)}{\omega + ipv_2 + p^2\left(D_2 + \left(G^{2}_{11}\right)^2
\left(\sqrt{2D_2(\omega + ipv_1 + D_2p^2/2)}\right)^{-1}\right)}.
\ee

For large times we assume the real-space scaling form
$S_2(x,t) = t^{-1/z} h\left(\frac{(x-v_2 t)^z}{t}\right)$
with dynamical exponent $z>1$. This is equivalent to the scaling forms
\begin{equation}
\hat{S}_2(p,t) = \rme^{- i v_2 p t} f(|p|^z t), \quad
\tilde{S}_2(p,\omega) = |p|^{-z} g \left(\frac{\omega + ipv_2}{|p|^{z}}\right)
\label{ScalingFormS2}
\end{equation}
for the Fourier- and Laplace transforms respectively. By introducing the shifted Laplace
parameter $\tilde{\omega} := \omega + ipv_2$ one finds that the
leading small-$p$ behaviour of the Laplace transform \eref{LPMCO} comes from the
term proportional to $v_1-v_2$ under the square root. This yields
$z=3/2$ and
we obtain in the limit $\tilde{\omega} \rightarrow 0$
(with scaling variable $\tilde{\omega}/|p|^z$ kept fixed) after performing
the inverse Laplace transformation
\begin{equation}
\hat{S}_2(p,t)=\frac{1}{\sqrt{2\pi}}
\exp{\left(- i v_2 p t -C_0 |p|^{3/2} t \left[ 1 - i \, \sign(p(v_1-v_2)) \right]\right)}
\label{S2(k,t)}
\end{equation}
with
\begin{equation}
C_0= \frac{\left(G^{2}_{11}\right)^2}{2\sqrt{ D_1 |v_2-v_1|}}.
\label{C0}
\end{equation}
We recognize here the characteristic function of an $\alpha$-stable L\'evy distribution
\bel{Def:Levy}
\hat{\phi} (p;\mu,c,\alpha,\beta) := \exp{\left( ip\mu - |cp|^{\alpha}
(1-i\beta\tan{\left(\frac{\pi \alpha}{2}\right)}\sign(p))\right)}
\ee
with $\mu = -  v_2  t$, $\alpha=3/2$, $c = (C_0 t)^{2/3}$
and maximal asymmetry $\beta=\sign(v_1-v_2)=\pm 1$.

We remark that in real space the asymmetric L\'evy scaling function has only one heavy tail decaying
as $1/x^{1+\alpha}$ which in a finite system leads to finite size corrections of order $1/L^{1+\alpha}$
for times $t \ll L^\alpha$.
The other tail, that extends away from the position of the other mode, decays exponentially.
This effect, which defines a kind of light cone, is a classical analogue of the Lieb-Robinson-bound
for the spreading of perturbations
in quantum systems \cite{Lieb72}.
The scaling function (\ref{S2(k,t)}) is similar to  the
one found to  describe the hydrodynamics of the anharmonic chain  in the case
of an "even potential", see \cite{Spoh14}.

\begin{figure}[htbp]
\begin{center}
 {\includegraphics[width=0.9\textwidth]{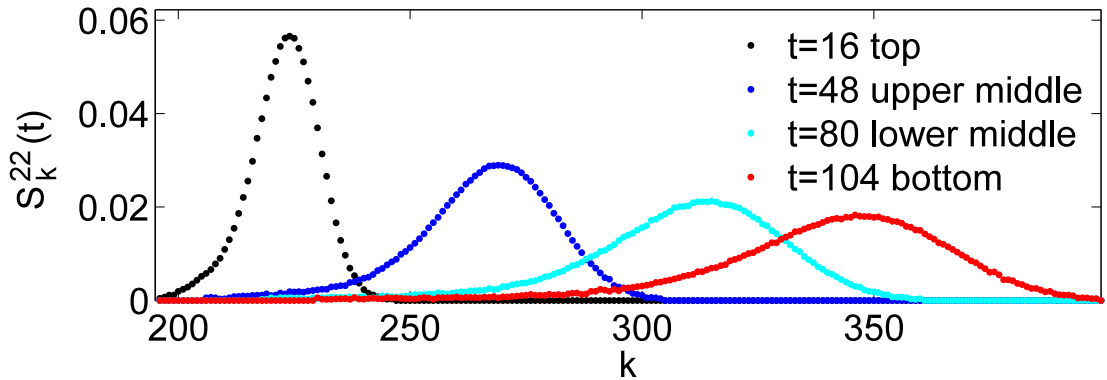}} 
\caption{(Colour online)  Dynamical structure function $S^{22}_k(t)$ for $3/2$-L\'evy mode with $v_2=1.3$
measured by Monte Carlo simulation
at different times, averaged over $18 \cdot 10^7$
histories. Parameters:  $L=600,\ \gamma=2.5,\ b = 1, \ \rho_1=0.2, \ \rho_2=0.2$.
  Statistical errors are smaller than symbol size.}%
\label{FigNewModePeaks}
\end{center}
\end{figure}

Monte-Carlo simulation data for the 3/2-L\'evy mode are shown in \Fref{FigNewModePeaks} for small times
up to $t\approx 100$. The mode moves with a velocity that, numerically, 
cannot be distinguished from the theoretical
prediction $v_2 = 1.3$. Indeed, one expects the error in the velocity, if at all, to be small,
since the velocity comes from mass conservation and is an exact constant 
for all times even on the lattice \cite{Popk03}.

\begin{figure}[ptb]
\centerline{
\includegraphics[width=0.7\textwidth]{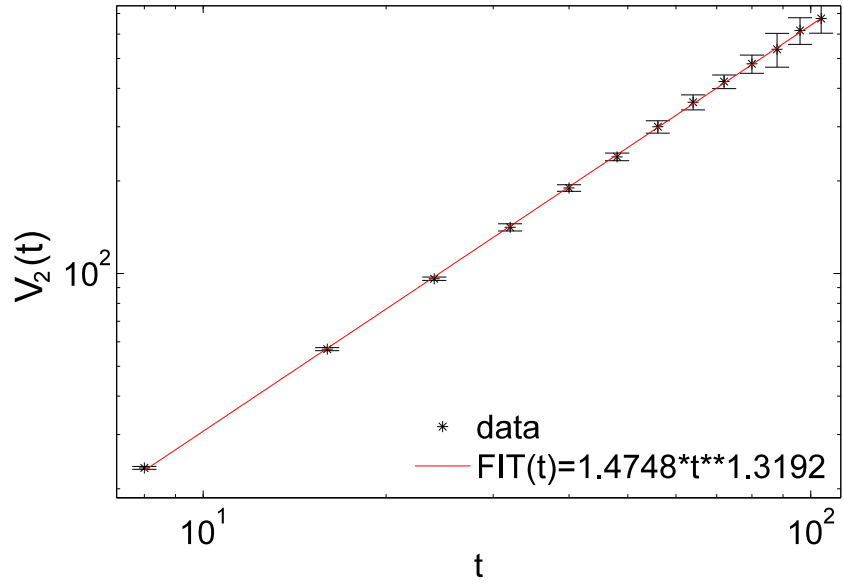}
}
\caption{(Color online) Variance $V_2(t)$ of the measured dynamical structure function shown in Fig. \ref{FigNewModePeaks}
versus time.
}
\label{Fig_NewModeVariance}
\end{figure}

\begin{figure}[htbp]
\begin{center}
 {\includegraphics[width=0.7\textwidth]{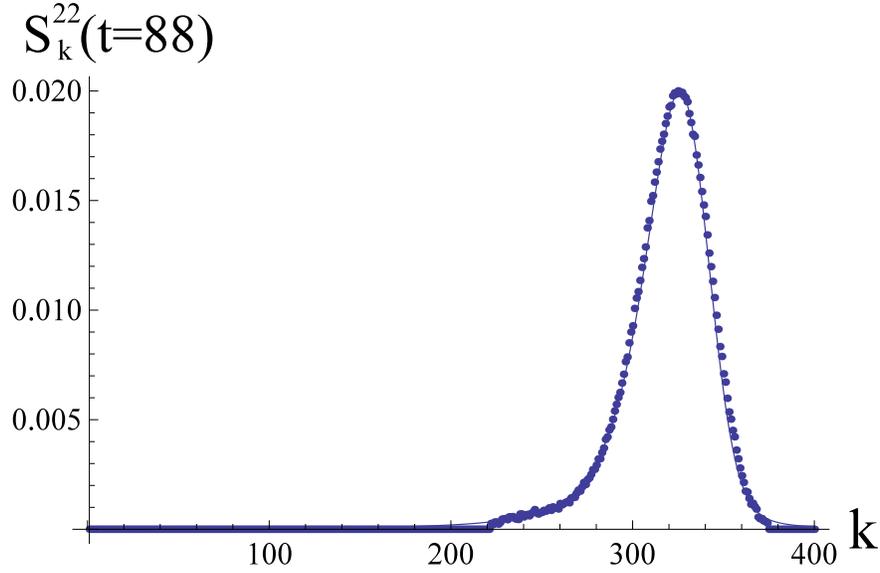}} 
\caption{(Colour online)  Fit of the dynamical structure function $S^{22}_k(t)$
for
 time $t=88$ with a 3/2-stable L\'evy distribution with asymmetry
 $\beta=-0.692$. For parameters see Fig. \ref{FigNewModePeaks}.}
\label{FigStableFitNewModePeak}
\end{center}
\end{figure}

The scaling exponent and asymmetry predicted by mode coupling theory are in a good agreement
with the Monte Carlo simulations, see Figs.~\ref{Fig_NewModeVariance},~\ref{FigStableFitNewModePeak}.
In \Fref{Fig_NewModeVariance} we show the growth of the variance $V_2(t)$
of the measured 3/2-L\'evy
mode. This quantity is not infinite for finite times, since the (single) heavy tail of the
asymptotic asymmetric L\'evy scaling function (\ref{S2(k,t)})
is cut off at finite times by the coupling to the other mode at a distance of the order $(v_2-v_1)t$.
Thus one expects
the empirical variance $V_2(t)$ to be finite but growing in time. Mass conservation together with dynamical
scaling predicts a growth $V_2(t) \propto t^{\nu}$ with $\nu=2/z$ \cite{Popk14}. The measured exponent
$\nu_{exp} \approx 1.32$ is very close to the theoretical value $\nu=4/3$ even for the
early time regime shown in the figures.\\

\begin{figure}[htbp]
\begin{center}
 {\includegraphics[width=0.7\textwidth]{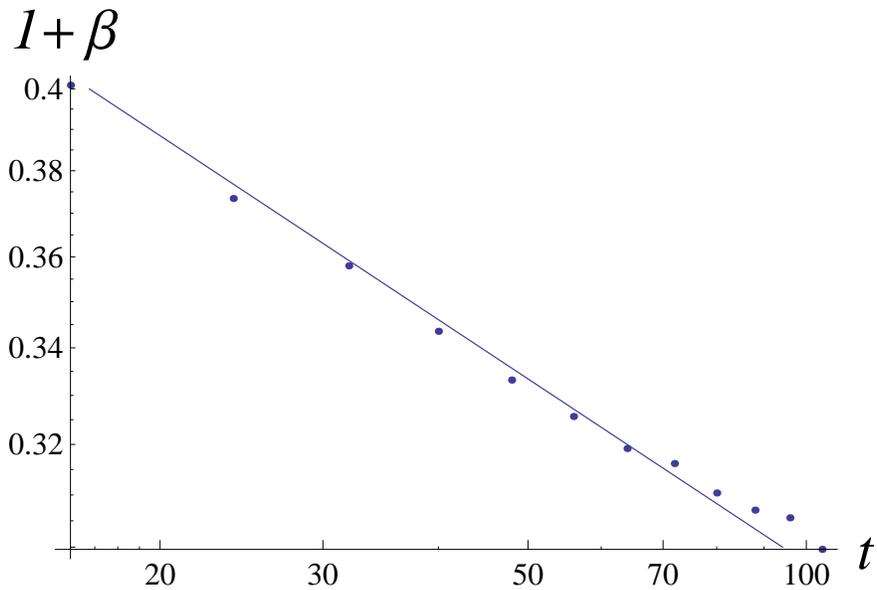}} 
\caption{(Colour online) Asymmetry $1+\beta$ versus time, obtained by fitting
the numerically obtained dynamical structure function with the PDF of $3/2$ L\'evy stable law.
The line with the power law $\propto t^{-1/6}$ is a guide to the eye. For
  parameters see \Fref{FigNewModePeaks}.
}%
\label{FigAsym}
\end{center}
\end{figure}

The only parameter that has slow convergence to the asymptotic value 
is the asymmetry of the scaling function. A similar phenomenon is
discussed in \cite{Spoh14} in terms of corrections to scaling of the memory kernel 
for the 5/3-L\'evy mode. They are shown to
vanish slowly with a power law decay in time.
Here we measure the deviation of the asymmetry from its asymptotic value. The
measured quantity $1+\beta_{exp}$ decreases monotonically with time.
The decay is approximately algebraic with exponent $\approx 1/6$, see Fig. \ref{FigAsym}.\\

\subsection{Two golden mean modes}

We consider now the case where both self-coupling coefficients 
$G^{\alpha}_{\alpha\alpha}$ of the mode coupling matrix
vanish and {\it both} subleading coefficients $G^{\alpha}_{\beta\beta}$ are non-zero 
and in general unequal.
In this case one cannot use the Gaussian or the KPZ scaling function as an input into the
mode coupling equations. However, the equations
give a self-consistency relation which allows one to compute the scaling function for the two
modes, see \cite{Spoh14b} for the symmetric case where $G^{1}_{22}=G^{2}_{11}$.
For the generic non-symmetric case $G^{1}_{22}\neq G^{2}_{11}$ the calculation of \cite{Spoh14b}
is not directly applicable. However, one can adopt a similar philosophy with two
scaling functions
\begin{equation}
\hat{S}_1(p,t) = \rme^{- i v_1 p t} g(b |p| t^\beta) , \quad
\hat{S}_2(p,t) = \rme^{- i v_2 p t} h(c |p|^\gamma t)
\label{ScalingFormGM}
\end{equation}
as input, which, in addition to the {\it a priori} unknown
dynamical exponents $\gamma$ and $1/\beta$, have different scale factors $b,c$ as free variables.
With this ansatz one obtains by power counting the consistency conditions 
$\gamma=1+\beta$ and $\gamma=1/\beta$
for the dynamical exponent.
From the mode coupling equations we have computed also the scale factors.
These computations are lengthy, but straightforward. 
With the relabelling $\hat{S}_{-}(p,t) \equiv \hat{S}_{1}(p,t)$,
$\hat{S}_{+}(p,t) \equiv \hat{S}_{2}(p,t)$
one arrives at
\bel{GMscalingfunction}
\hat{S}_{\pm}(p,t) = \frac{1}{\sqrt{2\pi}} \exp{\left(- i v_\pm pt -
C_\pm |p|^\varphi t \left[1 \pm i \sign{(p(v_- - v_+))}\tan{\left(\frac{\pi\varphi}{2}\right)}\right]\right)}
\ee
with golden mean $\gamma = \varphi \equiv (1+\sqrt{5})/2$ and the scale factors
\be
C_\pm = \half |v_+-v_-|^{1- \frac{2}{\varphi}}
\left(\frac{2G^{1}_{22}G^{2}_{11}}{\varphi \sin{\left(\frac{\pi\varphi}{2}\right)}}\right)^{\varphi-1}
\left(\frac{G^{1}_{22}}{G^{2}_{11}}\right)^{\pm(1+\varphi)}.
\ee
Notice that $\varphi-1=1/\varphi$.

For numerical simulation of this new universality class we choose the parameter manifold
\eref{GMparameters} of the two-lane model where one has the characteristic velocities
\be
v_\pm = (1+\gamma\rho_{2})(1-2\rho_{1}) \pm \gamma \sqrt{\rho_1(1-\rho_1)\rho_2(1-\rho_2)}.
\ee
We have chosen $\gamma=2.5,\, b=0.625$ and $\rho_1=0.25, \,\rho_2=0.2$
corresponding to the mode coupling matrices
\begin{equation}
G^{1}=\left(
\begin{array}{cc}
 0 & -0.406416 \\
 -0.406416 & -0.105726
\end{array}
\right), \quad
G^{2}=\left(
\begin{array}{cc}
 -0.812833 & -0.052863 \\
 -0.052863 & 0
\end{array}
\right)
\label{G2forGM}
\end{equation}
and transformation matrix
\begin{equation}
R^{-1}=\left(
\begin{array}{cc}
 - 0.734553 & 0.734553 \\
 0.678551 & 0.678551
\end{array}
\right).
\label{GMparametersEigenSystem}
\end{equation}
The columns of $R$ are the eigenmodes with velocities $v_1\equiv v_- = 0.3170$,
$v_2 \equiv v_+ = 1.183$ respectively.
In order to measure the dynamical exponent $\varphi\approx 1.618$, which is rather close to
$z=5/3\approx 1.667$ appearing in the
(5/3L,KPZ) scenario studied in \cite{Popk14}, we focus on the large-time regime rather than
looking into corrections to scaling for the asymmetry as done above. We use 
simulation parameters $n=300$, $m=3$, $\tau=120$ and $n=1000$, $m=46$, $\tau=120$.

\begin{figure}[htbp]
\begin{center}
 {\includegraphics[width=0.8\textwidth]{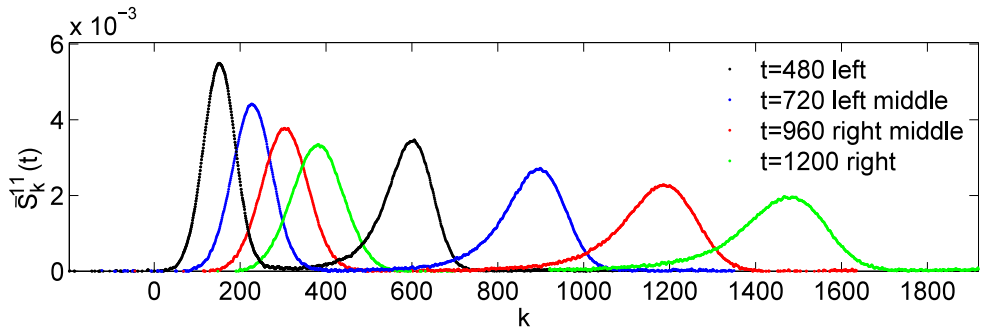}}
\caption{(Colour online)  Dynamical structure function showing both modes for
particles on chain 1, for the golden mean mode with $v_2=1.183$  at different times.
 Parameters:  $L=10^6,\gamma=2.5,b=0.625, \rho_1=0.25,\rho_2=0.2$.
  Statistical errors are smaller than symbol size.}%
\label{FigGMPeaks}
\end{center}
\end{figure}

Fig. \ref{FigGMPeaks} shows the measured dynamical structure function for both golden modes
moving on lattice 1.
The peaks are well separated already 
at the earliest time $t=480$ shown in the figure. 
The center of mass velocities have no perceptible deviation from the theoretical predictions.

\begin{figure}[ptb]
\centerline{
\includegraphics[width=0.7\textwidth]{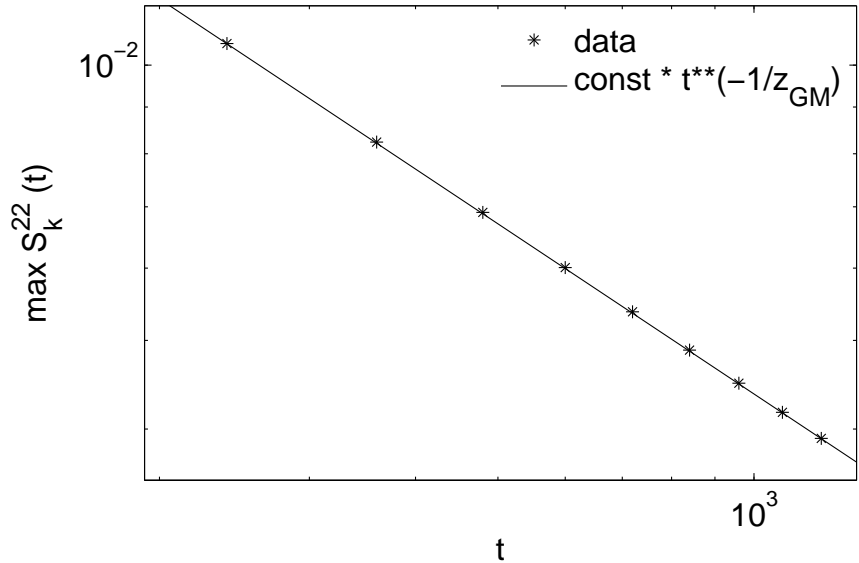}
}
\caption{(Color online) Maximum of the dynamical structure function for mode 2 
versus time, plotted in double logarithmic scale.
The line with the theoretically predicted slope (notice: $z_{GM}\equiv \varphi$ ) is a guide to the eye.
Model parameters are as in Fig.~\ref{FigGMPeaks}.
}
\label{FigGMmaximums}
\end{figure}

In Fig.~\ref{FigGMmaximums} we plot the maximum of the dynamic
structure function for mode $2$ (which scale as $t^{-1/z}$) as a function of time.
A least square fit with 95$\%$ confidence bounds gives a measured dynamical exponent 
$z=  1.619$, with error bars $1.613< z < 1.624$. This agrees 
well with the theoretically predicted golden mean value
 $z=\varphi \approx 1.618$.
 
\begin{figure}[ptb]
\centerline{
\includegraphics[width=0.7\textwidth]{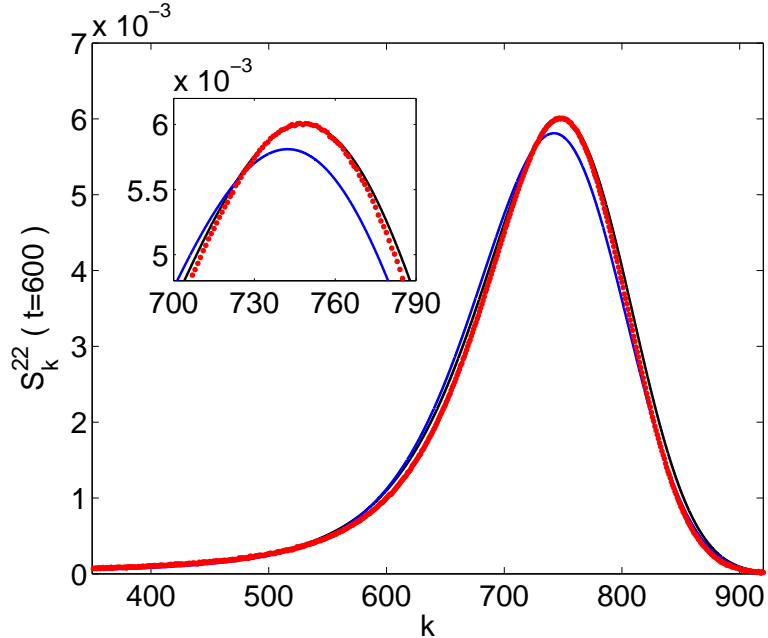}
}
\caption{(Color online) Measured dynamic structure function for mode $2$ at time $t=600$.
Monte Carlo data correspond to red dots, and black (blue) curves correspond to the
best least square fits of the Monte Carlo data
with the   $z=\varphi$ ($z=5/3$) stable L\'evy distribution with maximal asymmetry,
and theoretically predicted center of mass position. The inset shows a close-up view of the
peak region.
Model parameters are as in Fig.~\ref{FigGMPeaks}.
}
\label{FigGMLevyHeatfit}
\end{figure}

To investigate the convergence of the scaling function,
we plot both the measured structure function and the theoretically predicted $\varphi$-stable distribution
for a fixed time, see  Fig.~\ref{FigGMLevyHeatfit}. The theoretical prediction 
is well borne out by the simulation. Small deviations are visible in the right (fast decaying) tail, see
also the closeup view shown in the inset of Fig.~\ref{FigGMLevyHeatfit}.
A fit with a maximally asymmetric 5/3-stable L\'evy distribution shows a markedly poorer agreement.

Finally, we remark that the left peak in \Fref{FigGMPeaks} corresponding to mode $1$ is considerably less
asymmetric than the peak of mode $2$ shown in more detail in \Fref{FigGMLevyHeatfit}. To get some intuition
for this observation we point out to the numerical values $G^{1}_{22}$
and $G^{2}_{11}$ \eref{G2forGM}.
The ratio of their square is $(G^{1}_{22}/G^{2}_{11})^2\approx 0.017$, so the coupling 
strengths differ by almost two orders of magnitude.
If $G^{1}_{22}$ was zero, we would be back to the (D,$\frac{3}{2}$L) scenario discussed in the previous
subsection and mode 1 would be a symmetric Gaussian peak. Therefore one indeed expects
for mode $1$ at
finite times
a more symmetric function than predicted for the asymptotic regime.

\subsection{KPZ mode and $5/3$\,-\,L\'evy mode}

\begin{figure}[ptb]
\centerline{
\includegraphics[width=0.7\textwidth]{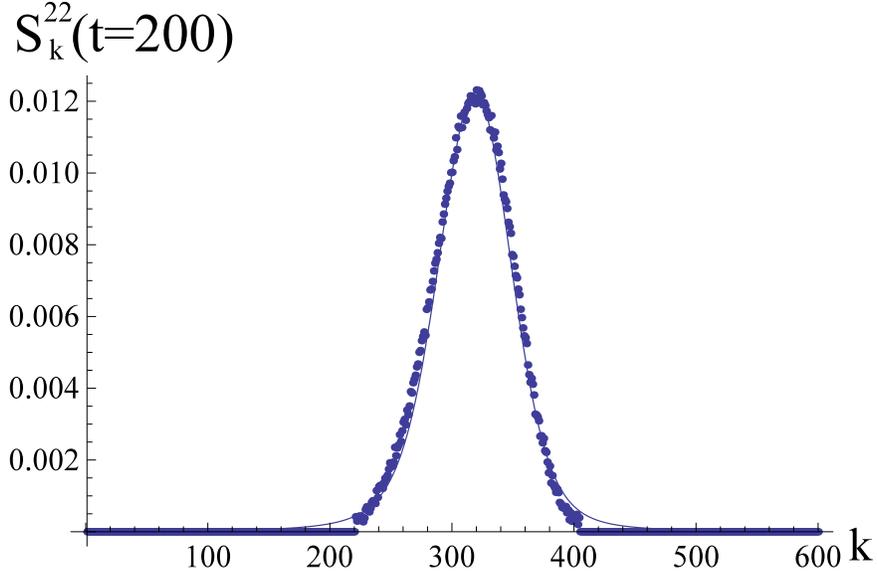}
}
\caption{(Color online) Measured dynamic structure function for mode 2 at time $t=200$. The curve
is a fit with the $5/3$-stable L\'evy distribution with non-maximal asymmetry..
}
\label{FigStableHeatfit}
\end{figure}

In \cite{Popk14} we reported the occurrence of the ($\frac{5}{3}$L,KPZ) universality class for the totally
asymmetric version of the two-lane exclusion process. The measured dynamical exponents were shown
to agree well with the theoretical prediction. Here we expand on these result by briefly discussing
the scaling function. In \Fref{FigStableHeatfit} one can see that a reasonable fit of the numerical
data can be obtained with a $5/3$-stable L\'evy distribution predicted by mode coupling theory
\cite{Spoh14}.\\

The measured dynamical structure function, however,
exhibits an asymmetry much less than the predicted maximal value. Indeed, for small times
its amplitude is rather small.
We attribute this discrepancy to finite-time effects, cf. the argument for the left GM mode
of the previous subsection. In order to substantiate this claim we show in Table
\ref{Tab:53asymmetry} numerically determined asymmetries. They grow in time, thus supporting the argument.
We do not have a theoretical prediction of how they should grow.

%

\begin{table}[htb]
  \begin{tabular}{c|ccccccccc}
  \hline \hline
\hspace{2mm} $t$  \hspace{2mm}  &\quad 20 &\quad 40 &\quad 60 &\quad 80 &\quad 100 &\quad 120 &\quad 140 &\quad 160 &\quad 200 \\
 $\beta$ &\quad -0.0229 &\quad -0.0504 &\quad -0.0685 &\quad -0.0797 &\quad -0.0825 &\quad -0.0872 &\quad -0.0918 &\quad -0.0916 &\quad -0.1000 \\
\hline \hline
  \end{tabular}
\caption{Asymmetry $\beta$ of a 5/3-L\'evy distribution obtained from a fit to the numerical data
for the 5/3-L\'evy mode of \cite{Popk14} for different early times $t \leq 200$. The predicted
asymptotic value is -1.
}
\label{Tab:53asymmetry}
\end{table}

\section{Conclusions}
\label{Sec:Conclusions}

We have studied time-dependent density fluctuations in driven diffusive system with two conservation
laws. For one conservation law it is well-established that the appropriate tool to describe the
universal properties of these fluctuations is non-linear fluctuating hydrodynamics \eref{normalmodes}.
Recent work, reviewed in \cite{Spoh14}, shows that the approach can be extended to
anharmonic chains with more than one conservation law and also to Hamiltonian dynamics
with three conservation laws \cite{vanB12}.
From the present work and our preliminary results reported in 
\cite{Popk14} we conclude that the predictions of the
theory apply also to driven diffusive systems with stochastic lattice gas dynamics with
two conservation laws. Specifically, for a
two-lane asymmetric simple exclusion process we argue that all theoretically
possible universality classes for two-component systems, discussed also in \cite{Spoh14b}, 
can be realized (see Table \ref{Tab:MCTscenarios}). Among these, our
Monte-Carlo simulations of a two-lane asymmetric exclusion process confirm two superdiffusive
universality classes which have gone unnoticed so far in the literature on driven diffusive systems.

Mode coupling theory not only predicts the dynamical exponents $z$ for these universality classes, but also
the scaling forms of the dynamical structure functions for these novel superdiffusive modes.
In most cases these scaling functions are $z$-stable L\'evy distributions with maximal asymmetry. The numerical
simulation  confirms these predictions with great
accuracy both for the $3/2$-mode and a golden mean mode with $z=(1+\sqrt{5})/2$ shown
to occur also in anharmonic chains \cite{Spoh14b}. For some modes the $z$-stable L\'evy distributions
provide excellent fits, but with an effective asymmetry that is not maximal. However, our data
show that the numerically fitted asymmetry increases with time in the cases we considered, thus
supporting the notion that asymptotically the maximal value will be reached.

Which universality classes actually occur in a system at given values of the physical parameters of the
model is completely encoded in the stationary
current-density relation $\vec{j}(\rho_1,\rho_2)$, no other knowledge about a given model is required.
The stationary compressibility
matrix $K(\rho_1,\rho_2)$, related to the current-density relation through a time-reversal symmetry proved in
\cite{Gris11}, allows for the prediction also of the scale factors that enter the scaling functions,
unless diffusive modes are relevant.
Thus generically the scaling functions are completely determined by two simple stationary properties:
The current-density relation $\vec{j}(\rho_1,\rho_2)$ and the compressibility matrix $K(\rho_1,\rho_2)$.
Going beyond specific lattice gas models, we have computed the mode coupling matrices in general form
for arbitrary input data, i.e., arbitrary current-density relation and compressibility matrix.
From the diagonal matrix elements of these one can then directly read off the scaling functions
for arbitrary two-component systems, except in the presence of the diffusive universality class where
the scale factors contain a phenomenological diffusion coefficient not predicted by the theory and 
which may modify the KPZ mode.

It is interesting to notice that all possible scenarios of universality classes
(see Table \ref{Tab:MCTscenarios}) can be realized with the simple current-density 
relation \eref{current}. This relation is minimal in the sense that the non-linearity of the conserved
current $j_\lambda$ is only quadratic and the coupling of this non-linearity to the other
conserved quantity is only linear, i.e., $\rho_\lambda^2\rho_\mu$ for $\lambda \neq \mu$. 
Thus it is not necessary to have a more complicated
current-density relation in order to observe all allowed universality classes.
Moreover, this minimal current-density relation has the nice property that one does
not expect logarithmic corrections to diffusive modes \cite{Devi92}.
Our two-lane exclusion process, which is an extension of the model studied by us
previously \cite{Popk14}, provides a simple microscopic realization for this minimal current-density
relation.

Throughout this discussion we have tacitly assumed that the current-density
relation is strictly hyperbolic, i.e., the collective velocities $v_\alpha$ of the two modes
are different. This assumption is crucial for the decoupling argument for the modes
that underlies the mode-coupling computations. Indeed,
the nonequilibrium time reversal symmetry \eref{timerev} \cite{Gris11} rules out umbilic
points (where $v_1=v_2$) in any model which has minimal current-density relation
and at the same time a diagonal compressibility matrix. Therefore in the
model presented here the issue does not actually arise. However,
umbilic points are a generic feature of more complicated models, either with the
same minimal current-density relation, but a non-diagonal compressibility matrix \cite{Popk12},
or for non-minimal current-density relations \cite{Erta92}. From numerical observations \cite{Erta92}
one expects the dynamical exponent $z=3/2$ as for KPZ, but non-KPZ scaling functions.
How mode coupling theory can predict the behaviour at umbilic points is an open problem.
It would also be interesting to extend mode coupling theory to predict
the convergence of the finite-time asymmetry in the L\'evy distribution to
the asymptotic maximal value.

\section*{Acknowledgements}

The authors are indebted to H. Spohn for pointing out the possibility of the
golden mean universality class prior to publication of \cite{Spoh14b}.
Indeed, this hint motivated us to systematically explore the structure of the mode-coupling matrices.
We also thank him and G. Stoltz for useful comments on a preliminary version of the manuscript and
M. Barma, R. Livi, H. Posch, A. Schadschneider and  H. van Beijeren for enlightening discussions.
Financial support by Deutsche Forschungsgemeinschaft (DFG) is gratefully acknowledged.
We thank the INFN and
the Galileo Galilei Institute for Theoretical Physics, where part of this work was done,
for hospitality and for partial support.

\appendix

\section{Mode coupling matrices for strictly hyperbolic two-component systems}
\label{Normalmodesbasis}

\subsection{Notation}

We consider a general system with two conservation laws. For definiteness we choose the
language of driven diffusive systems with currents $j_\lambda(\rho_1,\rho_2)$,
$\lambda=1,2$ for the conserved densities $\rho_\lambda$.
We define the general flux Jacobian
\be
J = \left( \ba{cc}  J_{11} &  J_{12} \\  J_{21} &  J_{22} \ea \right)
\ee
with matrix elements
\be
 J_{\lambda\mu} = \frac{\partial j_\lambda}{\partial \rho_\mu}
\ee
The transposed matrix is denoted $J^T$.

We define
\bel{def:discrim}
\delta := ( J_{11}- J_{22})\sqrt{1+\frac{4 J_{12} J_{21}}{( J_{11}- J_{22})^2}}
\ee
which is the signed square root of the discriminant of the characteristic
polynomial of $J$ with the sign given by $J_{11}- J_{22}$.
The two eigenvalues of $J$ are
\bel{charvel}
v_{\pm} = \frac{1}{2} \left( J_{11}+ J_{22} \pm \delta \right).
\ee
We associate velocity $v_-$ with eigenmode 1 and  $v_+$ with eigenmode 2,
irrespective of the sign of $v_--v_+$ which is equal to the sign of
$J_{22}- J_{11}$.

The matrix elements of the Hessians are denoted
\be
H^{\lambda} = \left( \ba{cc} h^{\lambda}_1 & \bar{h}^{\lambda} \\ \bar{h}^{\lambda} & h^{\lambda}_2 \ea \right)
\ee
with
\be
h^{\lambda}_1 = (\partial_1)^2 j_\lambda, \quad h^{\lambda}_2 = (\partial_2)^2 j_\lambda, \quad \bar{h}^{\lambda} = \partial_1 \partial_2 j_\lambda.
\ee
They are symmetric by definition.

The compressibility matrix is denoted
\be
K = \left( \ba{cc} \kappa_1 & \bar{\kappa} \\ \bar{\kappa} & \kappa_2 \ea \right)
\ee
It is symmetric by definition. Without loss of generality
we can assume $\kappa_1 \kappa_2 \neq 0$ since a vanishing self-compressibility
corresponds to a ``frozen'' lane without fluctuations which would reduce the
dynamics of the two-lane system to a dynamics with a single conservation law.
Time-reversal yields the Onsager-type symmetry \cite{Gris11}
\bel{timerev}
J K = K J^T
\ee
which implies
\bel{sym}
 J_{21} \kappa_1 -  J_{12} \kappa_2 = ( J_{11}- J_{22}) \bar{\kappa}.
\ee
Relation \eref{timerev} also guarantees that the eigenvalues $v_\pm$ of a physical flux
Jacobian $J$ are generally real. A related symmetry relation was noted earlier in the
context of classical fluids \cite{Spoh91}.

We point out the somewhat surprising fact that for {\it any} model with $\bar{\kappa}=0$, 
i.e., whenever the
stationary distribution factorizes in the conserved quantities,
the compressibilities satisfy
$ J_{21} \kappa_1 =  J_{12} \kappa_2$.
Thus a vanishing
cross derivative $ J_{\lambda\mu}$ for one of the currents implies a vanishing cross
derivative $J_{\mu\lambda}$ also of the other, without any {\it a priori} assumption
on the stochastic dynamics. The same is true also on parameter manifolds where
$J_{11}=J_{22}$.

\subsection{Normal modes}

We focus on the strictly hyperbolic case $v_+ \neq v_- $
corresponding to $\delta \neq 0$.
Since $J$ is not assumed to be symmetric we have to distinguish right (column) and left (row)
eigenvectors, denoted
by $\vec{c}^\pm$ and $\vec{r}^\pm$, respectively. Here
\be
\vec{c}^\pm = \left( \ba{c} c_1^\pm \\ c_2^\pm \ea \right), \quad \vec{r}^\pm = \left( r_1^\pm , \, r_2^\pm \right).
\ee
We normalize them to obtain a biorthogonal
basis with scalar product
\be \vec{r}^\sigma \cdot \vec{c}^{\sigma^\prime} := r_1^\sigma c_1^{\sigma^\prime} + r_2^\sigma c_2^{\sigma^\prime} =
\delta_{\sigma,{\sigma^\prime}}
\ee
with $\sigma,{\sigma^\prime} \in \{\pm\}$. Using
\be
\frac{ J_{22}- J_{11}-\delta}{2\sqrt{ J_{12} J_{21}}} =  \frac{2\sqrt{ J_{12} J_{21}}}{ J_{11}- J_{22}-\delta}
\ee
this yields
\bea
\vec{c}^\pm & = & \frac{1}{2 \delta y_\pm} \left( \ba{c} 2 J_{12} \\  J_{22}- J_{11} \pm \delta \ea \right), \\
\vec{r}^\pm & = &
\frac{y_\pm}{\delta \pm ( J_{22}- J_{11})} \left( 2 J_{21} , \,   J_{22}- J_{11} \pm \delta \right)
\eea
with arbitrary normalization constants $y_\pm$.

Next we introduce (bearing in mind that $\delta \neq 0$)
\bel{diagonalizer}
R = \left( \ba{cc} r_1^- & r_2^- \\ r_1^+ & r_2^+ \ea \right), \quad
R^{-1} = \left( \ba{cc} c_1^- & c_1^+ \\ c_2^- & c_2^+ \ea \right).
\ee
Biorthogonality and normalization give $R R^{-1} = 1$.
The fact that $R$ contains the left eigenvectors as its rows implies
$ RJ = \Lambda R$ where $\Lambda = diag(v_-,v_+)$. Therefore
\be
R J R^{-1} = \Lambda.
\ee
Then the linearized Eulerian hydrodynamic equations \eref{hyper} read
\be
\pdt \vec{\phi} + \Lambda \pdx \vec{\phi} = 0
\ee
with $\vec{\phi} = R \vec{u}$.

The diagonalizer $R$ is uniquely defined up to multiplication
by an invertible diagonal matrix which is
reflected in the arbitrariness of the normalization factors $y_\pm$.
In order to fix these constants we first observe that from \eref{timerev} it
follows that
$R (JK) R^T = \Lambda (RKR^T) = R(KJ^T) R^T = (RKR^T) \Lambda$.
Hence $RKR^T$ must be diagonal since $\Lambda$ is diagonal.
This allows us to fix the normalization constants $y_{\pm}$
by demanding
\bel{norm}
RKR^T = \mathds{1}.
\ee
This normalization
condition has its origin in the fact that the structure matrix $\bar{S}(k,t)$
(whose components are the dynamical structure functions \eref{dynstrucfun})
is by definition normalized such that $\sum_k \bar{S}(k,t) =K$, see next subsection.
For computing the normalization factors we first consider $ J_{12} J_{21}\neq 0$.\\

It is convenient to parametrize $R$ by diagonal matrices $Z=diag(z_-,z_+),\,U=diag(1,u)$
and an orthogonal matrix $O$ such that
\bel{R}
R = Z O U = \left( \ba{cc} z_- \cos{\vartheta} & -u z_-\sin{\vartheta} \\ z_+\sin{\vartheta} & uz_+ \cos{\vartheta} \ea \right)
\ee
with
\bel{def:phi_u}
\tan{\vartheta} = \frac{ J_{11}- J_{22} + \delta}{2\sqrt{ J_{12} J_{21}}}, \quad u = \sqrt{\frac{ J_{12}}{ J_{21}}}.
\ee
Notice that $ J_{12} J_{21}\neq 0$ implies $u\neq 0,\, \sin{\vartheta}\neq 0$
and $\cos{\vartheta}\neq 0$.
There are several useful identities involving the rotation angle $\vartheta$, viz.
$\tan{(2\vartheta)} = 2\sqrt{ J_{12} J_{21}}/( J_{22}- J_{11})$,
$\sqrt{ J_{12} J_{21}}(\cos^2{\vartheta}-\sin^2{\vartheta}) = ( J_{22}- J_{11}) \cos{\vartheta}\sin{\vartheta}$ and
$\delta = ( J_{22}- J_{11}) (\cos^2{\vartheta}-\sin^2{\vartheta}) + 4 \sqrt{ J_{12} J_{21}} \,
\cos{\vartheta}\sin{\vartheta} = ( J_{22}- J_{11}) \cos{(2\vartheta)} + 2 \sqrt{ J_{12} J_{21}} \, \sin{2\vartheta} =
( J_{22}- J_{11})/\cos{(2\vartheta)} = 2\sqrt{ J_{12} J_{21}}/\sin{(2\vartheta)}$.

Now we use that for $\bar{\kappa}\neq 0$ one can write
\be
U J U^{-1} =  \mu UKU +  \nu \mathds{1}
\ee
with
\be
\mu = \frac{ J_{21}}{\bar{\kappa}}, \quad \nu = \frac{1}{2}\left(  J_{11}+ J_{22} - \frac{ J_{21}\kappa_1+ J_{12}\kappa_2}{\bar{\kappa}}\right).
\ee
Therefore
\bea
RKR^T & = & ZOU K UO^TZ = \frac{1}{\mu} \left( ZOU J U^{-1}O^TZ - \nu Z^2\right) \nonumber \\
& = &  \frac{1}{\mu} \left(\ba{cc} v_- -z_-^2 \nu & 0 \\ 0 & v_+ - z_+^2 \nu \ea \right)
\eea
which yields
\bel{def:zpm}
z_\pm^{2} =  \frac{v_\pm -\mu}{\nu}.
\ee
By comparing with \eref{diagonalizer} one finds that
the normalization factors for the eigenvectors are given by $y_- =  u z_- \sin{\vartheta}$,
$y_+ =  u z_+ \cos{\vartheta}$.
For $\bar{\kappa}= 0$
one obtains directly from \eref{sym} and \eref{R} that
\bel{zpmspecial}
y_-^2 = \frac{\sin^2\vartheta}{\kappa_2}, \quad y_+^2 = \frac{\cos^2\vartheta}{\kappa_2}.
\ee

Even though not relevant for the two-lane model of this paper we mention for
completeness that
some care with limits has to be taken when $ J_{12} J_{21}=0$.
First notice that in this case the physical requirement $\kappa_1\kappa_2\neq 0$
implies $\bar{\kappa}\neq 0$.
Specifically for $ J_{12}=0$, $ J_{21}\neq 0$
one has $\delta =  J_{11}- J_{22}$, $v_- =  J_{11}$, $v_+ =  J_{22}$,
$ J_{21}\kappa_1 = ( J_{11}- J_{22})\bar{\kappa}$ and
\be
R = \left( \ba{cc}
\tilde{z}_- & 0
\\
\frac{\tilde{z}_+  J_{21}}{ J_{22}- J_{11}} & \tilde{z}_+
\ea \right)
\ee
with
\bel{def:tildezpm1}
\tilde{z}_-^{-2} = \kappa_1, \quad \tilde{z}_+^{-2} = \kappa_2 - \frac{ \bar{\kappa}^2}{ \kappa_1}.
\ee
Notice that here strict hyperbolicity implies $ J_{11}\neq  J_{22}$ so that $R$ is well-defined.

Similarly one obtains for $ J_{21}=0$, $ J_{12}\neq 0$ with $v_- =  J_{11} \neq v_+ =  J_{22}$
the relation $ J_{12}\kappa_2 = ( J_{22}- J_{11})\bar{\kappa}$ and
\be
R = \left( \ba{cc}
\hat{z}_- & \frac{\hat{z}_-  J_{12}}{ J_{11}- J_{22}}
\\
0 & \hat{z}_+
\ea \right)
\ee
with
\bel{def:tildezpm2}
\hat{z}_-^{-2} = \kappa_1 - \frac{ \bar{\kappa}^2}{ \kappa_2}, \quad \hat{z}_+^{-2} =  \kappa_2.
\ee
If $ J_{12}= J_{21}=0$ then $J$ is diagonal. For the strictly hyperbolic case $J_{11} \neq J_{22}$
one necessarily has $\bar{\kappa}=0$ and the normalization condition \eref{norm}
yields $R = diag(\kappa_1^{-1},\kappa_2^{-1})$.

\subsection{Normal modes and the microscopic dynamical structure function}

In order to explain the origin of the normalization condition and to apply it
to the two-lane model. We define the random variables $f^{\lambda}_k(t) := n^{(\lambda)}_k(t) - \rho_\lambda$ 
and
$f^{\lambda}_0 := f^{\lambda}_0(0)$ where the random variable $n^{(\lambda)}_k(t)$ is the particle number
on site $k$ of lane $\lambda$ with particle density $\rho_\lambda$
at time $t$. We also define the two-component column vector
$\vec{f}_k(t)$ with components $f^{\lambda}_k(t)$ and the two-component row vector
$\vec{f}_0^T := (f^{1}_0,\,f^{2}_0)$.
Expectation w.r.t. the stationary distribution is denoted
by $\exval{\cdot}$. By translation invariance and stationarity one has $\exval{f^{\lambda}_k(t)}=0$.
The expectation of a matrix is understood as the matrix
of the expectations of its components. Defining the $2\times 2$-matrix
$\bar{{\cal S}}_k(t) = \vec{f}_k(t) \otimes \vec{f}_0^T$ (the components of which
are random variables)
the dynamical structure matrix with components
\eref{dynstrucfun} can be written
$\bar{S}_k(t)= \exval{\bar{\cal S}_k(t)}$.

The normalization of the dynamical structure matrix, defined by the sum over the whole lattice,
is given by
\be
\sum_k \bar{S}_k(t) = K.
\ee
It is independent of time because of translation invariance and particle number conservation.
Now we consider the lattice normal modes
\be
\vec{\phi}_k(t) = R \vec{f}_k(t)
\ee
with components $\phi^{\alpha}_k(t)$
where $R$ is the diagonalizer \eref{diagonalizer}. In components
\be
\phi^{1}_k(t) = r_{11} f^{1}_k(t) + r_{12} f^{2}_k(t), \quad
\phi^{2}_k(t) = r_{21} f^{1}_k(t) + r_{22} f^{2}_k(t)
\ee
and similarly $\phi^{\alpha}_0 := \phi^{\alpha}_0(0)$.
In terms of the lattice normal modes
the structure matrix has the form $S_k(t) := \exval{{\cal S}_k(t)}$ with
${\cal S}_k(t) := \vec{\phi}_k(t) \otimes \vec{\phi}_0^T$. This yields
\be
S_k(t) = R \bar{S}_k(t) R^T
\ee
with matrix elements $S_{\alpha\beta}(k,t) = \exval{\phi^{\alpha}_k(t)\phi^{\beta}_0}$.
The desired normalization
\be
\sum_k S_k(t) = RKR^T=\mathds{1}
\ee
leads to the requirement \eref{norm}.

\subsection{Computation of the mode-coupling matrices}

The mode-coupling coefficients are given by
\bel{G}
G^{\gamma}_{\alpha\beta} := \frac{1}{2} \sum_{\lambda} R_{\gamma \lambda} \left[ (R^{-1})^T H^{\lambda} R^{-1}\right]_{\alpha\beta}.
\ee
where
$G^{\gamma}_{\alpha\beta}=G^{\gamma}_{\beta\alpha}$.
Using the previous results one finds for $G^{1}$ the matrix elements
\bea
\label{G111}
G^{1}_{11} & = & \frac{1}{2z_-} \left[  \cos^2{\vartheta}  \left(h^{1}_1 \cos{\vartheta}  - u h^{2}_1\sin{\vartheta} \right)
+ u^{-2} \sin^2{\vartheta} \left(h^{1}_2 \cos{\vartheta}  - u h^{2}_2 \sin{\vartheta} \right) \right. \nonumber \\
& & \left.
-2u^{-1}\cos{\vartheta}\sin{\vartheta} \left( \bar{h}^{1}\cos{\vartheta}  - u \bar{h}^{2} \sin{\vartheta} \right) \right]  \\
\label{G122}
G^{1}_{22} & = & \frac{z_-}{2z_+^2} \left[  \sin^2{\vartheta}  \left( h^{1}_1 \cos{\vartheta} - u h^{2}_1\sin{\vartheta} \right)   + u^{-2} \cos^2{\vartheta} \left( h^{1}_2 \cos{\vartheta}  - u h^{2}_2\sin{\vartheta} \right) \right. \nonumber \\
& & \left. + 2u^{-1}\cos{\vartheta}\sin{\vartheta} \left( \bar{h}^{1} \cos{\vartheta}  - u \bar{h}^{2}\sin{\vartheta}  \right) \right]   \\
\label{G112}
G^{1}_{12} & = &
\frac{1}{2z_+} \left[ \cos{\vartheta}\sin{\vartheta}   \left(  h^{1}_1 \cos{\vartheta}  - u h^{2}_1\sin{\vartheta}
- u^{-2} h^{1}_2 \cos{\vartheta}  + u^{-1} h^{2}_2\sin{\vartheta} \right) \right. \nonumber \\
& &  \left. u^{-1} (\cos^2{\vartheta}-\sin^2{\vartheta}) \left( \bar{h}^{1}\cos{\vartheta}  - u \bar{h}^{2}\sin{\vartheta}  \right)  \right] ,
\eea
and for $G^{2}$ one has
\bea
\label{G222}
G^{2}_{22} & = & \frac{1}{2z_+} \left[  \sin^2{\vartheta}  \left( h^{1}_1\sin{\vartheta}  +
u h^{2}_1\cos{\vartheta} \right) + u^{-2} \cos^2{\vartheta} (h^{1}_2 \sin{\vartheta}  + uh^{2}_2\cos{\vartheta} )
\right. \nonumber \\
& & \left. +2u^{-1}\cos{\vartheta}\sin{\vartheta} \left( \bar{h}^{1}\sin{\vartheta}
+ u\bar{h}^{2}\cos{\vartheta}  \right) \right] \\
\label{G211}
G^{2}_{11} & = & \frac{z_+}{2z_-^2}  \left[   \cos^2{\vartheta} \left( h^{1}_1\sin{\vartheta}  +
uh^{2}_1\cos{\vartheta}\right) + u^{-2} \sin^2{\vartheta} \left(h^{1}_2\sin{\vartheta}  + h^{2}_2 u\cos{\vartheta} \right) \right. \nonumber \\
& & \left. -2u^{-1}\cos{\vartheta}\sin{\vartheta} \left( \bar{h}^{1}\sin{\vartheta}
 + u\bar{h}^{2}\cos{\vartheta} \right) \right]  \\
\label{G212}
G^{2}_{12} & = & \frac{1}{2z_-} \left[ \cos{\vartheta}\sin{\vartheta}   \left(  h^{1}_1\sin{\vartheta}  +
u h^{2}_1\cos{\vartheta} - u^{-2} h^{1}_2\sin{\vartheta}  - u^{-1} h^{2}_2\cos{\vartheta} \right) \right. \nonumber \\
& & \left. u^{-1} (\cos^2{\vartheta}-\sin^2{\vartheta})
\left( \bar{h}^{1}\sin{\vartheta}  + u\bar{h}^{2}\cos{\vartheta}  \right) \right].
\eea
In terms of the model parameters $a,b,c,d,\kappa_{1,2},\bar{\kappa}$
The quantities $\vartheta$ and $u$ are given in \eref{def:phi_u} and the
quantities $z_\pm$ are given in \eref{def:zpm}. The parameter $\delta$
appearing in \eref{def:phi_u} is given in \eref{def:discrim}.

In order to analyze the manifolds where diagonal elements
of the mode coupling matrices vanish it is convenient
to introduce
\bel{g1}
g^{1}_1 := h^{1}_1, \, g^{1}_2 := u^{-2} h^{1}_2, \,\bar{g}^{1} :=u^{-1}\bar{h}^{1}
\ee
\bel{g2}
g^{2}_1 := u h^{2}_1, \, g^{2}_2 := u^{-1} h^{2}_2, \,\bar{g}^{2} :=\bar{h}^{2}.
\ee
and define the polynomials
\bea
\label{Ddiag11}
D^{1}_1(\omega) & := & g^{1}_1 - \left(g^{2}_1 + 2 \bar{g}^{1}\right) \omega
+ \left(g^{1}_2 + 2 \bar{g}^{2}\right) \omega^2 - g^{2}_2 \omega^3 \\
\label{Ddiag12}
D^{1}_2(\omega) & := & g^{1}_2 - \left(g^{2}_2 - 2 \bar{g}^{1}\right) \omega
+ \left(g^{1}_1 - 2 \bar{g}^{2}\right) \omega^2 - g^{2}_1 \omega^3 \\
\label{Ddiag21}
D^{2}_1(\omega) & := & g^{2}_1 + \left(g^{1}_1 - 2\bar{g}^{2}\right) \omega
+ \left(g^{2}_2 - 2 \bar{g}^{1}\right) \omega^2 + g^{1}_2 \omega^3 \\
\label{Ddiag22}
D^{2}_2(\omega) & := & g^{2}_2 + \left( g^{1}_2 + 2 \bar{g}^{2}\right) \omega
+ \left( g^{2}_1 + 2 \bar{g}^{1}\right) \omega^2 + g^{1}_1 \omega^3.
\eea
with $\omega := \tan{\vartheta}$. Only the Hessian and the parameters $u$ and $\tan{\vartheta}$
given in \eref{def:phi_u} enter
these functions. They do not depend on the compressibilities.
Then one has
\bea
\label{Gdiag11}
G^{1}_{11} & = &  \frac{\cos^3\vartheta}{2z_-} D^{1}_1(\omega), \quad
G^{1}_{22} =  \frac{z_- \cos^3\vartheta}{2z^2_+} D^{1}_2(\omega) \\
\label{Gdiag22}
G^{2}_{11} & = &  \frac{z_+ \cos^3\vartheta}{2z^2_-} D^{2}_1(\omega), \quad
G^{2}_{22} =  \frac{\cos^3\vartheta}{2z_+} D^{2}_2(\omega).
\eea
Notice the symmetry properties $D^{1}_1(\omega) = -\omega^3 D^{2}_2(-\omega^{-1})$ and
$D^{1}_2(\omega) = - \omega^3 D^{2}_1(-\omega^{-1})$.

\end{document}